\documentclass[12pt,preprint,usenatbib]{emulateapj}

\usepackage{amssymb}
\usepackage{times}
\usepackage{pifont} 

\newcommand{\teff}{$T_{\mathrm{eff}}$}

\newcommand{\numax}{$\nu_{\mathrm{max}}$}
\newcommand{\dnu}{$\Delta\nu$}

\newcommand{\kepler}{\textit{Kepler}}
\newcommand{\keplers}{\textit{Kepler's}}
\newcommand{\kic}{\textit{KIC}}
\newcommand{\keplermission}{\textit{Kepler Mission}}

\shorttitle{An asteroseismic membership study of NGC~6791, NGC~6819, and NGC~6811}
\shortauthors{Stello et al.}

\begin{document}

\title{An asteroseismic membership study of the red giants in three open
  clusters observed by \kepler: NGC~6791, NGC~6819, and NGC~6811}  

\author{
Dennis~Stello,\altaffilmark{1} 
S{\o}ren~Meibom,\altaffilmark{2}
Ronald~L.~Gilliland,\altaffilmark{3}  %no feedback on lastest draft
Frank~Grundahl,\altaffilmark{4}
Saskia~Hekker,\altaffilmark{5}
Beno\^it~Mosser,\altaffilmark{6}   
Thomas~Kallinger,\altaffilmark{7,8} %no feedback on lastest draft
Savita~Mathur,\altaffilmark{9}
Rafael~A.~Garc{\'\i}a,\altaffilmark{10}
Daniel~Huber,\altaffilmark{1}
Sarbani~Basu,\altaffilmark{11}
Timothy~R.~Bedding,\altaffilmark{1}
Karsten~Brogaard,\altaffilmark{12,4} 
William~J.~Chaplin,\altaffilmark{5}
Yvonne~P.~Elsworth,\altaffilmark{5}
Joanna~Molenda-\.Zakowicz,\altaffilmark{13}
Robert~Szab\'o,\altaffilmark{14}
Martin~Still,\altaffilmark{15} %no feedback
Jon~M.~Jenkins,\altaffilmark{16} %no feedback
%William~J.~Borucki,\altaffilmark{26}
%David~Koch,\altaffilmark{26}
J{\o}rgen~Christensen-Dalsgaard,\altaffilmark{4} 
Hans~Kjeldsen,\altaffilmark{4} 
Aldo~M.~Serenelli,\altaffilmark{17} %no feedback
Bill~Wohler,\altaffilmark{18}
}
\altaffiltext{1}{Sydney Institute for Astronomy (SIfA), School of Physics, University of Sydney, NSW 2006, Australia}
\altaffiltext{2}{Harvard-Smithsonian Center for Astrophysics, 60 Garden Street, Cambridge, MA, 02138, USA}
\altaffiltext{3}{Space Telescope Science Institute, 3700 San Martin Drive, Baltimore, Maryland 21218, USA}
\altaffiltext{4}{Department of Physics and Astronomy, Aarhus University, Ny Munkegade 120, 8000 Aarhus C, Denmark}
\altaffiltext{5}{School of Physics and Astronomy, University of Birmingham, Edgbaston, Birmingham B15 2TT, UK}
\altaffiltext{6}{LESIA, CNRS, Universit\'e Pierre et Marie Curie, Universit\'e Denis Diderot, Observatoire de Paris, 92195 Meudon, France}
\altaffiltext{7}{Institute for Astronomy, University of Vienna, T \"urkenschanzstrasse 17, 1180 Vienna, Austria}
\altaffiltext{8}{Department of Physics and Astronomy, University of British Columbia, 6224 Agricultural Road, Vancouver, BC V6T 1Z1, Canada}
\altaffiltext{9}{High Altitude Observatory, NCAR, P.O. Box 3000, Boulder, CO 80307, USA}
\altaffiltext{10}{Laboratoire AIM, CEA/DSM-CNRS, Universit\'e Paris 7 Diderot, IRFU/SAp, Centre de Saclay, 91191, Gif-sur-Yvette, France}
\altaffiltext{11}{Department of Astronomy, Yale University, P.O. Box 208101, New Haven, CT 06520-8101}
\altaffiltext{12}{Department of Physics \& Astronomy, University of Victoria, P.O. Box 3055, Victoria, B.C., V8W 3P6, Canada}
\altaffiltext{13}{Instytut Astronomiczny Uniwersytetu Wroc{\l}awskiego, ul.\ Kopernika 11,51-622 Wroc{\l}aw, Poland}
\altaffiltext{14}{Konkoly Observatory of the Hungarian Academy of Sciences, Konkoly Thege Mikl\'os \'ut 15-17, H-1121 Budapest, Hungary}
\altaffiltext{15}{Bay Area Environmental Research Institute/NASA Ames Research Center, Moffett Field, CA 94035, USA}
\altaffiltext{16}{SETI Institute/NASA Ames Research Center, MS 244-30, Moffat Field, CA 94035, USA}
%\altaffiltext{10}{Indian Institute of Astrophysics, Koramangala, Bangalore 560034, India}
\altaffiltext{17}{Instituto de Ciencias del Espacio (CSIC-IEEC), Facultad de Ci\`encies, Campus UAB, 08193 Bellaterra, Spain}
%\altaffiltext{26}{NASA Ames Research Center, MS 244-30, Moffat Field, CA 94035, USA}
\altaffiltext{18}{Orbital Sciences Corporation/NASA Ames Research Center, Moffett Field, CA 94035}

\clearpage

\begin{abstract}
Studying star clusters offers significant advances in stellar
astrophysics due to the combined power of having many stars with essentially
the same distance, age, and initial composition.  This makes clusters excellent
test benches for verification of stellar evolution theory.  To fully exploit
this potential, it is vital that the star sample is 
uncontaminated by stars that are not members of the cluster.
Techniques for determining cluster membership therefore play
a key role in the investigation of clusters. 
We present results on three clusters in the \kepler~field of view based on
a newly established technique that uses asteroseismology to identify 
fore- or background stars in the field, which demonstrates advantages over
classical methods such as kinematic and photometry measurements.
Four previously identified seismic non-members in NGC~6819 are confirmed in
this study, and three additional non-members are found -- two in
NGC~6819 and one in NGC~6791.
We further highlight which stars are, or might be, affected by blending, which
needs to be taken into account when analysing these \kepler~data.
\end{abstract}


\clearpage

\section{Introduction} 
Determination of cluster membership is a crucial step in
the analysis of stellar clusters.  Stars in an open cluster are thought to have
formed from the same interstellar cloud of gas and dust, and hence share
a common age and space velocity. % making them an ideal for verifying
%stellar evolution theory. 
Cluster membership can therefore be inferred from the location of the stars along an
isochrone in the color-magnitude diagram (photometric membership), and from
their common space velocity (kinematic membership) measured as the 
line-of-sight radial velocity and the perpendicular proper motion. 
Recently, a new independent method was introduced by \citet{Stello10} who
performed an asteroseismic analysis based on the first month of data from the
\keplermission~\citep{Koch10} to infer the cluster membership for a small
sample of red giant stars in NGC~6819.   
Asteroseismology has the advantage that the oscillations in a star, which depend 
on the physical properties of the star's interior \citep{Dalsgaard04},
are independent of stellar distance, interstellar extinction, and
any random alignment between the space velocity of the cluster and field stars.
In particular, the so-called average large frequency separation, \dnu, between
consecutive overtone oscillation modes depends on the mean density of the star, and the
frequency of maximum oscillation power, \numax, is related to its surface
gravity and effective temperature.
Both \dnu~and \numax~
are known to scale with
%follow known scaling relations expressed by 
the basic stellar properties, $M$, $L$, and \teff,
\citep{Ulrich86,KjeldsenBedding95} and can therefore 
be used to infer those properties without relying on detailed modelling
of stellar interiors \citep[e.g.~][]{Stello08,Kallinger10}.  

We now have \kepler~time series photometry that span 10 times longer
than in the work by \citet{Stello10}.  In this paper we
are therefore able to present an asteroseismic membership analysis of a more
comprehensive set of red giant stars in three open clusters
within \kepler's fixed field of view: NGC~6791, NGC~6819, and NGC~6811. We are
further able to measure and make use of both \dnu~and \numax~for this purpose.
In addition, to facilitate our inference on cluster membership and
obtain more robust results we
include an investigation of \teff~found from different color indices and
present a detailed analysis of blending.  Our
membership results are compared with those from classical techniques.

We refer to our four companion papers for additional asteroseismic
exploration of the same cluster data including (i) the determination of
stellar mass and radius and cluster distances \citep{Basu11}, (ii) verification
of scaling relations for \dnu~and \numax~\citep{Hekker10a}, (iii)
derivation of a new scaling relation for oscillation amplitude
\citep{Stello11}, (iv) and mass loss properties of  
red giants during their transition between the hydrogen-shell and
core-helium-burning phases \citep{Miglio11}.  Like this paper, these
studies are based on the global asteroseismic seismic properties,
\dnu, \numax~and amplitude, while more detailed frequency analyses requires 
more data for these relative faint and crowded cluster stars.

\section{Target selection}\label{selection}
For the purpose of determining cluster membership we use only stars showing
oscillations that are stochatically driven by near-surface convection
(solar-like oscillations) 
because their seismic observables are strongly 
linked to the fundamental stellar properties described by well established
scaling relations (see Sect.~\ref{membership}). 
This limits our current study to the red giants as we
require the oscillations to be sufficiently sampled by the spacecraft's
half-hour cadence.

\begin{figure}
\includegraphics[width=8.cm]{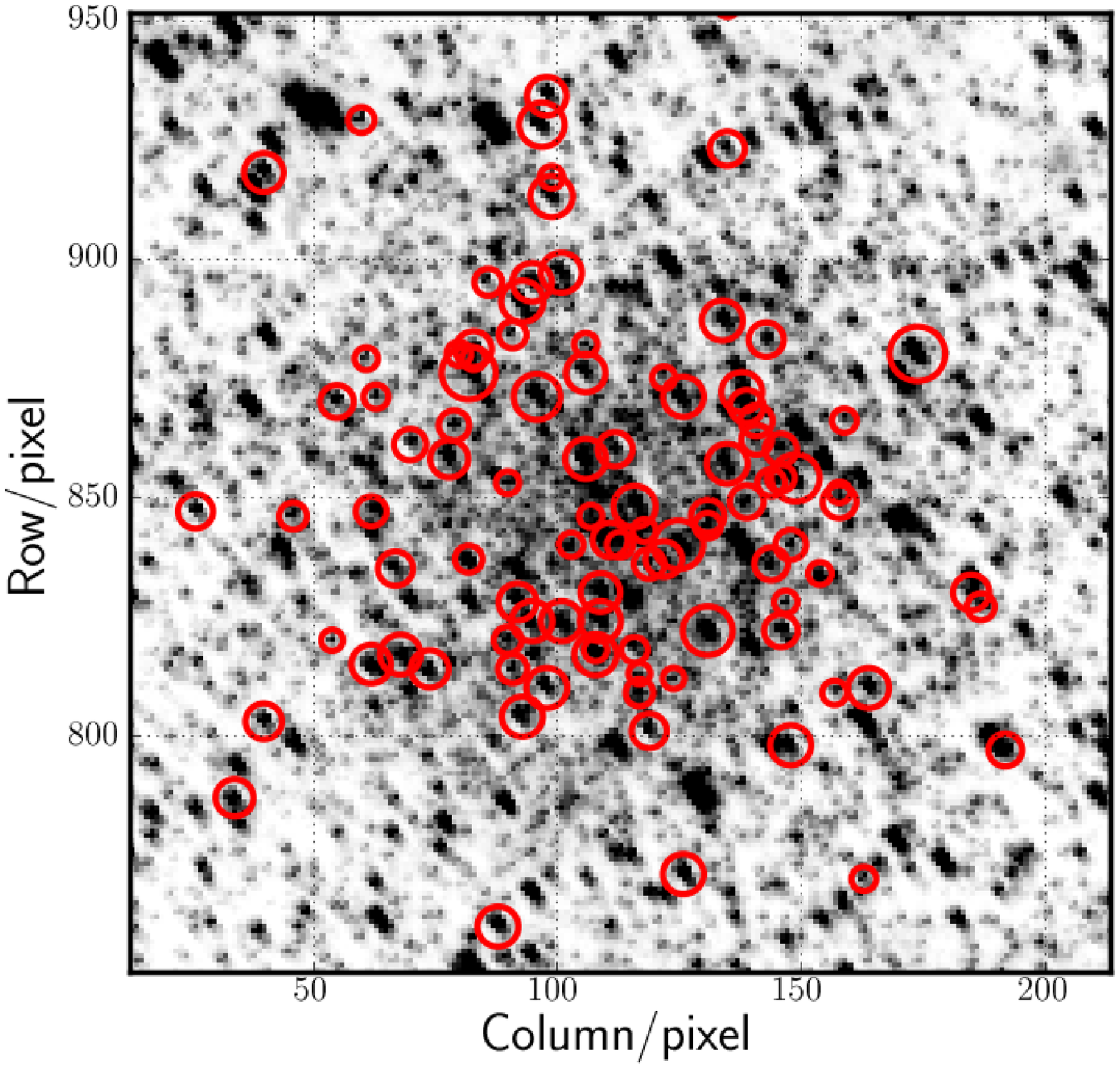}
\includegraphics[width=8.cm]{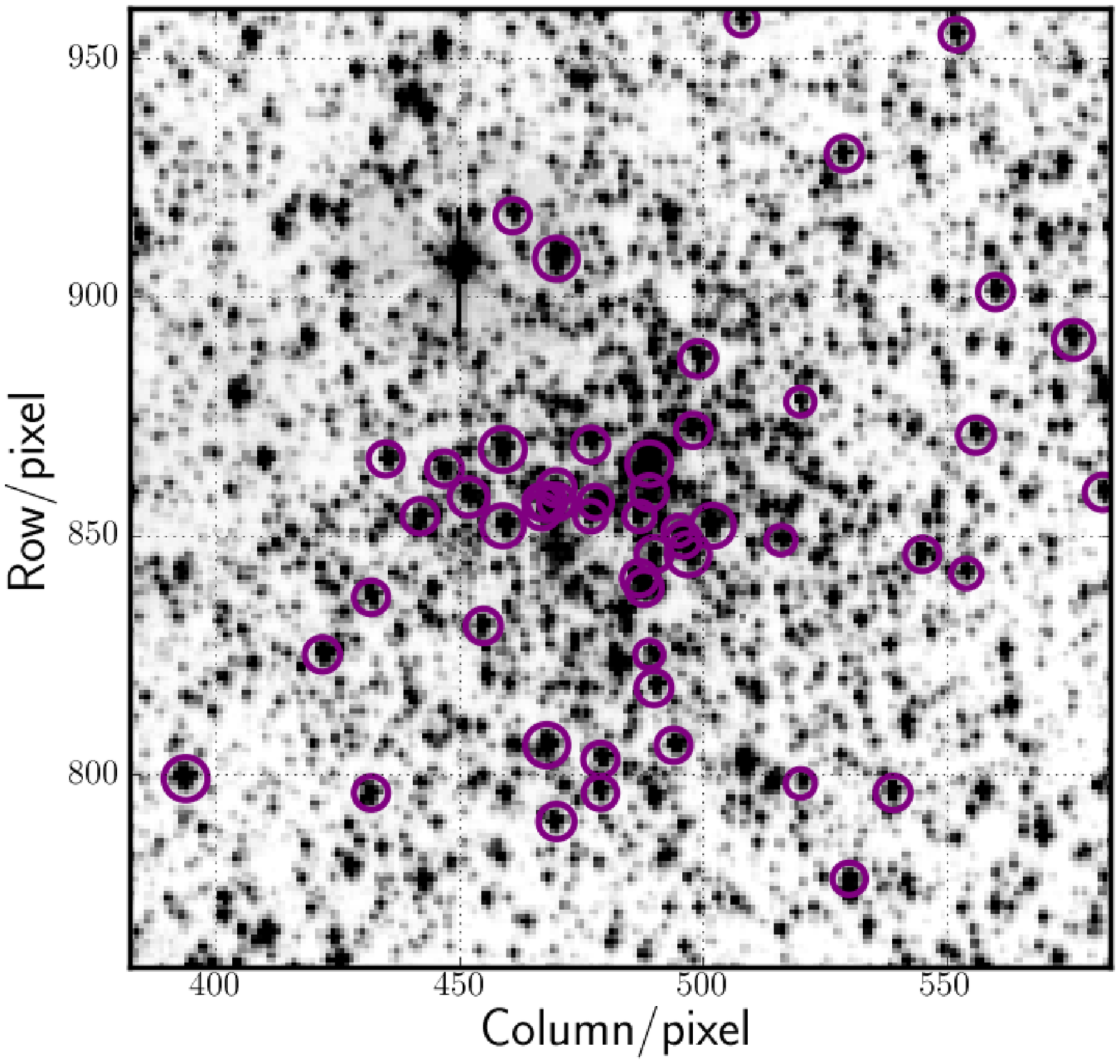}
\includegraphics[width=8.cm]{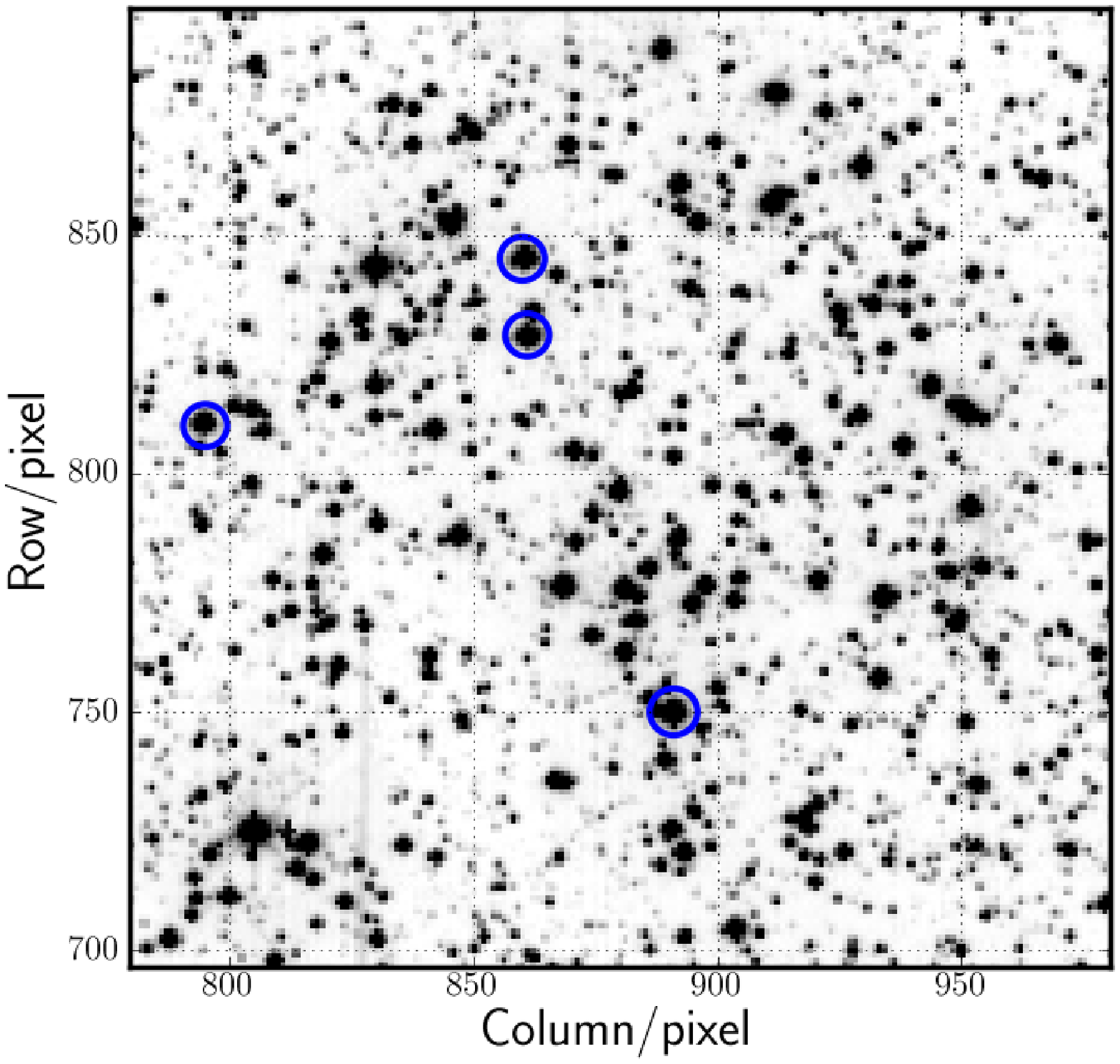}
%\plotone{f1.eps}
\caption{Cluster super stamps of NGC~6791 (top) and NGC~6819 (middle). A
  similar size stamp is shown for NGC~6811 (bottom) to illustrate the
  difference in crowding.  A total of 99 (NGC~6791), 54 (NGC~6819), and 4 
  (NGC~6811) of the pre-launch selected targets fall within these stamps
  (encircled). 
\label{finding}} 
\end{figure}

Large 200-pixel `super' stamps of the CCD images ($13'\!.3$ on the side)
centered on NGC~6791 and NGC~6819 are obtained at a half-hour cadence
throughout the mission, eventually providing photometric time
series (light curves) of all resolved stars within them
(Figure~\ref{finding}).  
However, these stamps require special image processing, which is
still pending.  We therefore base our current study on our initial
selection of individual cluster 
stars made prior to launch, which have followed the standard
\kepler~data reduction of the raw images (Sect.~\ref{observations}), and
includes a few stars outside the super stamps. 
Due to general limits on the number of stars that can be recorded by
\kepler~at any given time, the selection was aimed at maximising the
number of cluster members in our %target 
sample.

\begin{figure}
\includegraphics{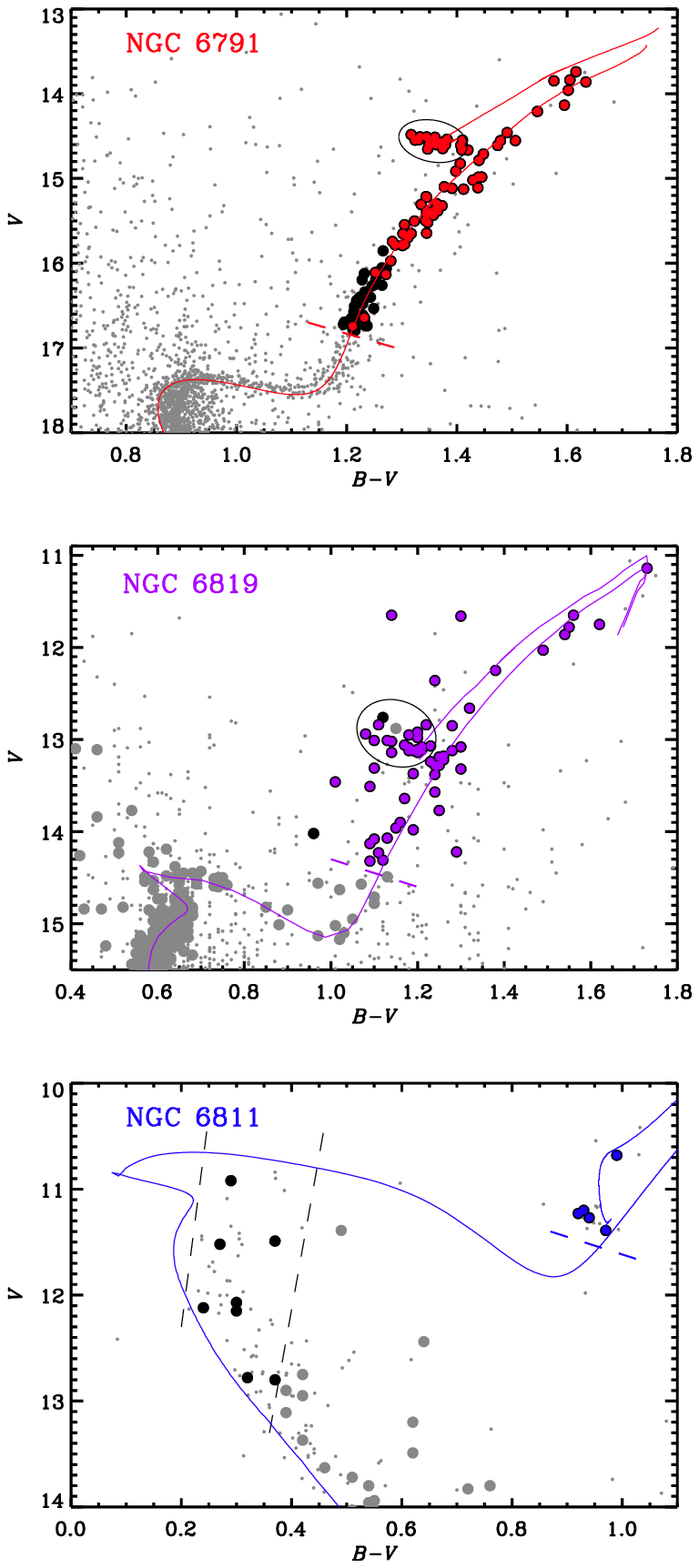}
%\plotone{f1.eps}
\caption{
  Color-magnitude diagrams of the clusters. Photometry is from
  \citet{Stetson03} (NGC~6791), \citet{Hole09} (NGC~6819), and the Webda
  database (http://www.univie.ac.at/webda/) (NGC~6811).  Representative
  isochrones from \citet{Pietrinferni04} 
  (NGC~6791 and NGC~6811) and \citet{Marigo08} (NGC~6819) are matched to
  the red giant stars to guide the eye.  Near horizontal dashed lines mark the
  sampling limit for solar-like oscillations with \keplers~long-cadence mode.
  Large black dots show stars for 
  which \kepler~data are currently available, while large colored dots
  (red, purple and blue) indicate the subset of these which show
  evidence of solar-like oscillations.  All large dots in the two
  lower panels show likely members from radial velocity surveys
  \citep[][~and unpublished work by Meibom]{Hole09}.  Vertical dashed lines mark the
  approximate location of the classical instability strip. Stars marked as
  'Clump star' in Tables~\ref{tab1} and~\ref{tab2} are encircled.
\label{cmd}} 
\end{figure}

For NGC~6791 no comprehensive kinematic membership study was available
so the selection was based on photometric membership which we determined using
the photometry by \citet{Stetson03}.  Only stars quite close to the
empirical isochrone in the color-magnitude diagram were chosen (Figure~\ref{cmd}, large dots). 
We note that with this strict selection criterion we risk missing genuine members that
are further away from the main cluster sequence, and hence might not sample
the full intrinsic scatter of the population.
Our selection provided 101 red giant stars. % (large dots in Figure~\ref{cmd}).

We selected 63 red giants in the open cluster NGC~6819 that had more than
80\% membership probability from the radial velocity survey of
\cite{Hole09}.  Being purely kinematic, this selection is more likely to
include stars that do not follow the standard single-star evolution.
Indeed, Figure~\ref{cmd} shows that a number of stars marked as kinematic members
(large dots) are quite far from the
empirical isochrone formed by the majority of stars.  

In the case of NGC~6811 we chose all stars determined to be possible
members from a preliminary radial velocity survey (Meibom)
(Figure~\ref{cmd}), which gave us five red giants in total.  
%There are no red giants selected in the fourth \kepler~cluster, NGC~6866,
%which is quite sparse and is omitted in this study.

For this purpose the data are obtained in the spacecraft's long-cadence mode .

\section{Observations and data reduction}\label{observations}
The photometric time series data presented here were obtained in 'long cadence'
($\Delta t \sim
30\,$min, \citet{Jenkins10a}) between 2009 May 12 and 2010 March 20, known as observing quarters
1--4 (Q1--Q4).  Within this period the spacecraft's long-cadence mode provided
approximately 14,000 data points per star.  The raw images were processed by 
the standard \kepler~Science Pipeline and included steps to remove
signatures in the data from sources such as pointing drifts, focus changes,
and thermal variations all performed during the Pre-search Data Conditioning
(PDC) procedure \citep{Jenkins10}.  
PDC also corrects for flux from neighboring stars 
within each photometric aperture based on a static aperture model.  However, this
model is not adequate for all stars, due to small changes in the
telescope point-spread-function and pointing between subsequent 
quarterly rolls when the spacecraft is rotated 90 degrees to align its
solar panels.  As a result the light curves show jumps
in the average flux level from one quarter to the next.  To correct for that we
shifted the average flux levels for each quarter to match that of the
raw (pre-PDC) data before stitching together the time series from all four
quarters. This ensured that the relative flux variations were consistent from
one quarter to the next. 
We compared our corrected (post-PDC) data with the raw data and also 
after we performed a number of ``manual'' corrections based
entirely on the appearance of the light curves (hence not taking
auxiliary house-keeping data such as pointing into account).  These
corrections included removal of 
outliers, jumps, and slow trends in a similar way as the approach by
\citet{Garcia11}.  The comparison revealed that for a few stars PDC did
not perform well, in which case we chose the raw or ``manually'' corrected
raw data.

\section{Blending and light curve verification}
The super stamps in Figure~\ref{finding} clearly illustrate that
blending is an issue we need to address before proceeding with the analysis of
these cluster data.  Some stars show clear signatures of blending arising from
the relatively large pixel scale ($\sim4''$) of the \kepler~photometer
compared to the fairly crowded cluster fields. 
Blending will give rise to additional light in the photometric aperture,
which will reduce the relative stellar variability, and
increase the photon counting noise.  In
severe cases, the detected stellar variability arises from
a blending star and not the target.

\begin{figure}
\includegraphics{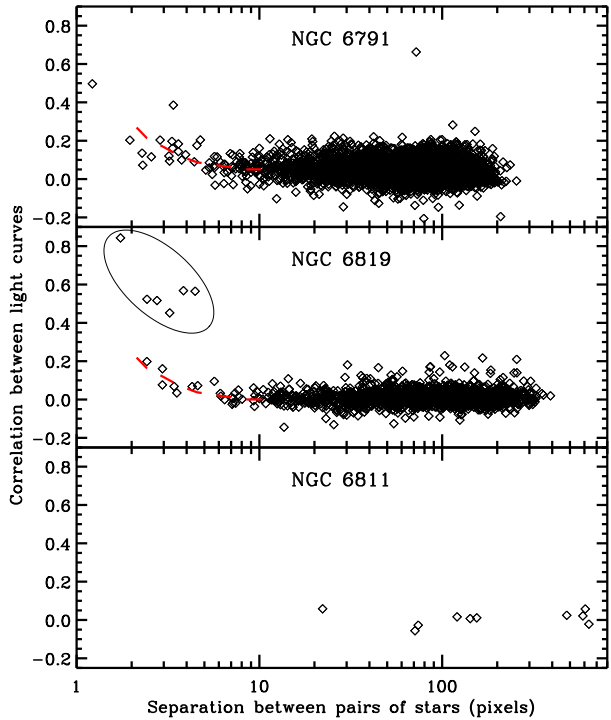}
%\plotone{f1.eps}
\caption{
  Correlation coefficients for light curves of all pairs of stars in our
  target list shown as a function of their separation on the CCD.  Fiducial
  dashed line indicates the average increase in correlation towards lower
  separations. There is one pair of stars in NGC~6791 that show large
  correlation despite their large separation (see text).  The six pairs of stars
  with abnormally high correlation coefficients in NGC~6819 are all blended
  by the W UMa variable KIC5112741 (see text).  
\label{corr}} 
\end{figure}

We have studied the effects from blending by
looking at correlations between light curves of all the target
stars (black, red, purple, and blue dots in Figure~\ref{cmd}).  
The light curve correlations show no significant increase unless
the stars are within approximately five pixels of each other 
and the blending star is at least as bright as the target (Figure~\ref{corr}). 
%The correlation plot indicates that increased correlation between star
%5024312 and 5024297 is due blending.  We get similarly conclusion for the
%pair 5024601 and 5024582.  However, the correlations of these two pairs are
%not particularly high compared to what we see between stars separated by
%up to hundreds of pixels, which are most likely dominated by
%correlations from instrumental or data reduction noise (Jon,Ron comments?). 
We visually assessed the light curves of all stars separated by
less than five pixels, and identified those that showed clear correlation
over extended periods of time as blends.  We list the blending stars in 
%The targets identified as blended from the derived correlation
%coefficients and visual inspection of the light curves are marked in
Table~\ref{tab1} (column-3), Table~\ref{tab2} (column-4), and
Table~\ref{tab5} (column-4).  
Also listed here, is the variability type of blending stars
identified from single light curves that clearly showed variability from
two stars.  If the variability included the expected seismic signal of the
target, under the assumption that the target was a cluster member, we
interpreted the additional signal as caused by a blend.
We see no blending for our 
NGC~6811 targets.  %Only stars for which stellar oscillations were detected
%(Sect.~\ref{extract}) are listed as targets.
We note that the stars for which we currently have light curves are
far from all stars in the vicinity of the clusters (see
Figure~\ref{finding}).  It is therefore likely that our correlation
analysis has not revealed all blends.  For the two most crowded clusters
(NGC~6791 and NGC~6819) there are about 
1000 stars brighter than $Kp\simeq16.5$ within the super stamps.  We expect
%to have light curves from 
a significant fraction of those stars to be accessible
when these stamps have 
been analyzed, which will further aid the characterisation of blending.

\begin{table*}
{\scriptsize%\footnotesize
\begin{center}
\caption{Target properties of NGC~6791.\label{tab1}}
\begin{tabular}{ccccccccc}
\tableline\tableline
Target ID$^a$ & Target ID     & Blend & Blend     & Clump & Membership  & Membership  & Member  & Seismic    \\ 
(KIC)         & (Stetson)$^b$ & known & potential & star  & (M\&P)$^c$  & (Garnavich)$^d$& (Other)$^e$  & member  \\ 
\multicolumn{1}{c}{[1]}&[2]   & [3]   & [4]       & [5]   & [6]         & [7]         &         & [8]        \\ 
\tableline		  				     	       				 
2297384 &        5583     &    No   &  Yes      &   Yes  &            &            &     &  Yes          \\ 			    
2297793 &       11539     &    No   &  No       &   No   &            &  R18/No?   &     &  No           \\ 
2297825 &       11957     &    No   &  No       &   Yes  &            &            &     &  Yes          \\ 			    
2435987 &         611     &    No   &  Yes      &   No   &            &            &     &  Yes          \\ 			    
2436097 &        1110     &    No   &  No       &   No   &            &            &     &  Yes          \\ 			    
2436209 &        1705     &    No   &  No       &   No   &            &            &     &  Yes          \\ 			    
2436332 &        2309     &    No   &  Yes      &   No   &  33\%      &            &     &  Yes          \\ 				    
2436417 &        2723     &    No   &  No       &   Yes  &  17\%      &            &  W  &  Yes          \\ 			    
2436458 &        2915     &    No   &  Yes      &   No   &  49\%      &            &     &  Yes          \\ 			    
2436540 &        3354     &    No   &  No       &   No   &            &            &     &  Yes          \\ 			    
2436593 &        3609     &    No   &  Yes      &   No   &  13\%      &            &     &  Yes          \\ 			    
2436676 &        4122     &    No   &  No       &   No   &  88\%      &            &     &  Yes          \\ 			    
2436688 &        4202     &    No   &  Yes      &   No   &  98\%      &            &     &  Yes          \\ 			    
2436732 &        4482     &    No   &  No       &   Yes  &  98\%      &            &     &  Yes          \\ 			    
2436759 &        4616     &    No   &  Yes      &   No   &  88\%      &            &     &  Yes          \\ 			    
2436814 &        4952     &    No   &  Yes      &   No   &            &            &  W  &  Yes          \\ 			    
2436818 &        4968     &    No   &  No       &   No   &            &            &     &  Yes          \\ 			    
2436824 &        4994     &    No   &  Yes      &   No   &  82\%      &            &     &  Yes          \\ 			    
2436900 &        5454     &    No   &  Yes      &   No   &  98\%      &            &     &  Yes          \\ 				    
2436912 &        5503     &    No   &  Yes      &   Yes  &  88\%      &            &     &  Yes          \\ 			    
2436944 &        5712     &    No   &  No       &   Yes  &  23\%      &            &     &  Yes          \\ 
2436954 &        5787     & 2436944 &  Yes      &   No   &            &            &     &  Yes          \\ 
2437040 &        6288     &    No   &  Yes      &   No   &  29\%      &            &     &  Yes          \\ 
2437103 &        6626     &    No   &  Yes      &   No   &            &            &     &  Yes          \\ 
2437171 &        6963     & 2437209 &  Yes      &   No   &  98\%      &  R4/Yes    &  O  &   ?           \\ 
2437240 &        7347     &    No   &  Yes      &   No   &  99\%      &            &     &  Yes          \\ 
2437270 &        7564     &    No   &  Yes      &   No   &  89\%      &            &     &  Yes          \\ 
2437325 &        7912     &    No   &  Yes      &   No   &  97\%      &            &     &  Yes          \\
2437340 &        7972     &    No   &  No       &   No   &  92\%      &  R19/Yes   &  C  &  Yes          \\ 				    
2437353 &        8082     &    No   &  Yes      &   Yes  &  93\%      &            & C,Gr&  Yes          \\ 			    
2437394 &        8317     &    No   &  Yes      &   No   &  98\%      &            &     &  Yes          \\ 				    
2437402 &        8351     &    No   &  Yes      &   No   &  96\%      &            &     &  Yes          \\ 			    
2437444 &        8563     &    No   &  No       &   No   &            &            &  C  &  Yes          \\ 				    
2437488 &        8865     & Binary* &  Yes      &   No   &            &            &     &  Yes          \\ 			    
2437496 &        8904     &    No   &  Yes      &   No   &  42\%      &  R12/Yes   &  O  &  Yes          \\ 			       
%Target ID$^a$ & Target ID      & Blend & Blend     & Clump & Membership  & Membership  & Seismic \\	                                   
%(KIC)         & (Stetson)$^b$  & known & potential & star  & (M\&P)$^c$  & (Garvanich)$^d$  & member  \\	   
%\multicolumn{1}{c}{[1]}&[2]    & [3]   & [4]       & [5]   & [6]         & [7]         & [8]     \\	   
2437507   &       8988 &    No   &  No       &   No  & 29\%  &            &  C & Yes     \\			   
2437564   &       9316 &    No   &  Yes      &   Yes & 24\%  &            & Gr & Yes     \\			   
2437589   &       9462 &    No   &  No       &   Yes & 99\%  &            &    & Yes     \\			   
2437653   &       9827 &    No   &  Yes      &   No  & 94\%  &            &    & Yes     \\			   
2437698   &      10135 &    No   &  No       &   Yes & 77\%  &            &    & Yes     \\			   
2437781   &      10674 &    No   &  Yes      &   No  &       &            &    & Yes     \\			   
2437804   &      10809 &    No   &  Yes      &   Yes & 97\%  &            &    & Yes     \\			   
2437805   &      10806 &    No   &  Yes      &   Yes & 97\%  &            & Gr & Yes     \\			   
2437816   &      10898 &    No   &  No       &   No  & 68\%  &            &  C & Yes     \\			   
2437851   &      11116 &    No   &  Yes      &   No  & 17\%  &            &    & ?       \\			   
2437933   &      11598 &    No   &  Yes      &   No  & 94\%  &            &    & Yes     \\			   
2437957   &      11797 &    No   &  Yes      &   No  & 11\%  &            &    & Yes     \\			   
2437965   &      11814 &    No   &  Yes      &   No  & 98\%  &  R8/Yes    &  C & Yes     \\			   
2437972   &      11862 &    No   &  Yes      &   No  & 85\%  &            &    & Yes     \\			   
2437976   &      11895 &    No   &  Yes      &   No  & 98\%  &            &    & Yes     \\			   
2437987   &      11938 &    No   &  Yes      &   Yes & 96\%  &            &    & Yes     \\			   
2438038   &      12249 &    No   &  Yes      &   No  & 96\%  &            &    & Yes     \\			   
2438051   &      12333 &    No   &  Yes      &   Yes & 99\%  &            &    & Yes     \\			   
2438140   &      12836 &    No   &  No       &   No  &       &            &    & Yes     \\			   
2438333   &      13847 &    No   &  No       &   No  &       &            &    & Yes     \\			   
2438421   &      14379 &    No   &  No       &   No  & 47\%  &  R7/Yes    &  O & ?       \\			   
2568916   &        996 &    No   &  No       &   Yes &       &            &    & Yes     \\			   
2569055   &       1904 &    No   &  No       &   Yes &       &            &    & Yes     \\			   
2569360   &       3754 &    No   &  No       &   No  & 94\%  &            &  W & Yes     \\			   
2569488   &       4715 &    No   &  No       &   Yes & 49\%  &            &  C & Yes     \\			   
2569618   &       5796 &    No   &  No       &   No  & 99\%  &            &    & Yes     \\			   
2569935   &       8266 &High peak*& No       &   No  &       &  R16/Yes   & O,C& Yes     \\			   
2569945   &       8395 &    No   &  Yes      &   Yes & 89\%  &            &    & Yes     \\			   
2570094   &       9786 &    No   &  Yes      &   No  & 85\%  &            &    & Yes     \\			   
2570172   &      10407 &    No   &  Yes      &   No  &       &            &    & Yes     \\			   
2570214   &      10695 &    No   &  Yes      &   Yes & 88\%  &            &    & Yes     \\			   
2570244   &      11006 &    No   &  Yes      &   No  & 90\%  &            &    & Yes     \\			   
2570384   &      12265 & Binary* &  Yes      &   No  &       &            &    & Yes     \\			   
2570518   &      13260 &    No   &  Yes      &   No  &       &            &    & Yes     \\			   
%          &            &         &           &       &       &            &      \\                           
\tableline
\end{tabular}
\tablenotetext{a}{Only targets for which we detect oscillations are listed.}
\tablenotetext{b}{IDs from \citet{Stetson03}.}
\tablenotetext{c}{Preliminary membership probabilities from radial velocity (Meibom \& Platais, priv. comm. (2010)).}
\tablenotetext{d}{IDs and membership from radial velocity \citep{Garnavich94}.}
\tablenotetext{e}{Members accoding to both radial velicity and [Fe/H]:
  W=\citet{WortheyJowett03}, O=\citet{Origlia06}, C=\citet{Carraro06}, Gr=\citet{Gratton06}.}
\tablenotetext{*}{Signal removed from light curve.}
%\tablenotetext{b}{Classifications from radial velocity measurements by
%  \citep{Hole09} (SM: single member; BM: binary member; BLM: binary likely member; U: Unknown).}
%\tablenotetext{c}{Membership probability from proper motion \citep{Sanders72}.}
\end{center}}
\end{table*}

%KIC     Member* Reference						 
%2436417 Y       W							 
%2436814 Y       W							 
%2437171 Y       O							 
%2437340 Y       C							 
%2437353 Y       C,Gr						 
%2437444 Y       C							 
%2437496 Y       O							 
%2437507 Y       C							 
%2437564 Y       Gr							 
%2437589 Y       Ga	???						 
%2437805 Y       Gr							 
%2437816 Y       C							 
%2437965 Y       C							 
%2438421 Y       O							 
%2569360 Y       W							 
%2569488 Y       C							 
%2569935 Y       O,C						 
%*Membership is based on radial velocity in Ga, and on both radial ve locity and [Fe/H] in other	    
%references.							 
%									 
%References:							 
%C       Carraro et al. 2006, ApJ 643, 1151				 
%Ga      Garnavich et al. 1994, ApJ 107, 1097			 
%Gr      Gratton et al. 2006, ApJ 642, 462				 
%O       Origlia et al. 2006, ApJ 646, 504				 
%W       Worthey & Jowett 2003, PASP 115, 96                         

%\input{tab3_delux}
\begin{table*}
{\footnotesize
\begin{center}
\caption{Target properties of NGC~6819.\label{tab2}}
\begin{tabular}{ccccccccccc}
\tableline\tableline
Target ID$^a$ & Target ID         & Target ID     & Blend   & Blend     & Clump  & Class.            &  Membership        & Membership     & Photometric & Seismic   \\
(KIC)         & (Hole et al.)$^b$ & (Sanders)$^c$ & known   & potential &star    & (Hole et al.)$^b$ &(Hole et al.)$^b$ &(Sanders)$^c$ & member      & member       \\
\multicolumn{1}{c}{[1]} & [2] & [3]   & [4]     & [5]   & [6]    &  [7]    & [8]         &  [9]        & [10]       & [11]    \\
\tableline		  	  	    		                       
4936335&       007021  &     9   &     No       &  No   &  No    &    SM    & 95\%  &  68\%   &  Yes    &   No     \\
4937011&       007017  &    90   &     No       &  No   &  No    &    SM    & 95\%  &  90\%   &  Yes    &   No     \\
4937056&       002012  &   103   &     No       &  No   &  Yes   &    BM    & 95\%  &  92\%   &  Yes    &  Yes     \\
4937257&       009015  &   144   &     No       &  No   &  No    &    SM    & 88\%  &  80\%   &   No    &   No     \\
4937576&       005016  &   173   &     No       &  No   &  No    &    SM    & 91\%  &  88\%   &  Yes    &   Yes    \\
4937770&       009024  &         &    High peak*&  No   &  No    &    SM    & 94\%  &         &   No    &   Yes    \\
4937775&       009026  &         &     No       &  No   &  No    &    BM    & 91\%  &         &   No    &   Yes    \\
5023732&       005014  &    27   &     No       &  No   &  No    &    SM    & 94\%  &  90\%   &   Yes   &   Yes    \\
5023845&       008010  &    36   &     No       &  No   &  No    &    SM    & 95\%  &  89\%   &   Yes   &   Yes    \\
5023889&       004014  &    42   &     No       &  No   &  No    &    U     & 95\%  &  90\%   &   No    &   No     \\
5023931&       007009  &    43   &     No       &  No   &  No    &    BM    & 84\%  &  91\%   &   Yes   &   Yes    \\
5023953&       003011  &    45   &     No       &  No   &  Yes   &    BLM   &       &  90\%   &   Yes   &   Yes    \\
5024043&       008013  &    58   &     No       &  No   &  Yes   &    SM    & 95\%  &  65\%   &   Yes   &   Yes    \\
5024143&       007005  &    65   &     No       &  No   &  No    &    SM    & 94\%  &  69\%   &   Yes   &   Yes    \\
5024240&       008007  &         &     No       &  No   &  No    &    BM    & 88\%  &         &   Yes   &   Yes    \\
5024268&       002003  &    78   &$\delta\,$Scuti*& No  &  No    &    SM    & 93\%  &  92\%   &   No    &   No     \\
5024272&       003003  &    79   &     No       &  No   &  No    &    SM    & 95\%  &         &   No    &   No     \\
5024297&       008003  &    87   &      5024312 &  Yes  &  No    &    SM    & 89\%  &  92\%   &   Yes   &   Yes    \\
5024312&       013002  &    86   &      5024297 &  Yes  &  No    &    SM    & 89\%  &  87\%   &   Yes   &   Yes    \\
5024327&       011002  &    96   &     No       &  No   &  Yes   &    SM    & 94\%  &  88\%   &   Yes   &   Yes    \\
5024404&       003004  &    98   &     No       &  No   &  No    &    SM    & 93\%  &  81\%   &   Yes   &   Yes    \\
5024405&       004001  &   100   &     No       &  Yes  &  No    &    SM    & 93\%  &  91\%   &   Yes   &   Yes    \\
5024414&       006002  &   106   &     No       &  Yes  &  Yes   &    SM    & 95\%  &  91\%   &   Yes   &   ?      \\
5024456&       001002  &   110   &     M-giant* &  Yes  &  No    &   SM     & 88\%  &  72\%   &   Yes   &   Yes    \\
5024476&       001006  &   111   &     No       &  No   &  Yes   &   BLM    &       &  89\%   &   Yes   &   ?      \\
5024512&       003001  &   116   &     No       &  Yes  &  No    &   SM     & 93\%  &  90\%   &   Yes   &   Yes    \\
5024517&       002001  &         &     No       &  Yes  &  No    &   SM     & 88\%  &         &   Yes   &   ?      \\
5024582&       009002  &   118   &     5112741* &  Yes  &  Yes   &   BLM    &       &  87\%   &   Yes   &   Yes    \\
       &               &         &     5024601  &       &        &          &       &         &         &          \\
5024583&       007003  &   119   &     Binary*  &  Yes  &  No    &   SM     & 95\%  &  92\%   &   Yes   &   Yes    \\  
5024601&       004002  &   124   &     5024582  &  Yes  &  Yes   &   SM     & 92\%  &  86\%   &   Yes   &   Yes    \\  
5024750&       001004  &   141   &     No       &  No   &  No    &   SM     & 93\%  &  83\%   &   Yes   &   Yes    \\ 
5024851&       002008  &   152   &     No       &  No   &  No    &   BLM    &       &  64\%   &   Yes   &   Yes    \\ 
5024967&       006009  &   158   &     No       &  No   &  Yes   &   SM     & 92\%  &  87\%   &   Yes   &   Yes    \\ 
5111718&       008018  &    10   &     No       &  No   &  No    &   SM     & 95\%  &  91\%   &   Yes   &   Yes    \\
5111940&       005012  &    28   &     No       &  Yes  &  No    &   SM     & 94\%  &  79\%   &   Yes   &   Yes    \\ 
5111949&       004011  &    30   &     No       &  No   &  Yes   &   SM     & 93\%  &  83\%   &   Yes   &   Yes    \\ 
5112072&       009010  &    39   &     No       &  No   &  No    &   SM     & 95\%  &  91\%   &   Yes   &   Yes    \\ 
5112288&       002007  &    64   &     No       &  No   &  Yes   &   SM     & 93\%  &  90\%   &   Yes   &   Yes    \\ 
5112361&       004008  &    70   &     No       &  No   &  No    &   BM     & 91\%  &  78\%   &   No    &   Yes    \\ 
5112373&       005005  &    74   &     No       &  No   &  Yes   &   SM     & 95\%  &  87\%   &   Yes   &   Yes    \\ 
5112387&       003007  &    73   &     No       &  No   &  Yes   &   SM     & 95\%  &  88\%   &   Yes   &   Yes    \\ 
5112401&       003009  &    75   &     No       &  No   &  Yes   &   SM     & 95\%  &  92\%   &   Yes   &   Yes    \\ 
5112403&       005004  &    77   &     No       &  No   &  No    &   SM     & 91\%  &  89\%   &   Yes   &   Yes    \\ 
5112467&       006003  &    85   &     No       &  Yes  &  Yes   &   SM     & 95\%  &  87\%   &   Yes   &   Yes    \\ 
5112481&       001007  &    93   &     No       &  No   &  No    &   SM     & 92\%  &  89\%   &   Yes   &   Yes    \\ 
5112491&       010002  &    89   &     No       &  Yes  &  Yes   &   SM     & 95\%  &  92\%   &   Yes   &   Yes    \\ 
5112730&       004005  &   128   &     No       &  No   &  Yes   &   SM     & 93\%  &  62\%   &   Yes   &   Yes    \\  
5112734&       012002  &   130   &     5112741* &  Yes  &  No    &   SM     & 91\%  &  90\%   &   Yes   &   Yes    \\
5112744&       005011  &   127   &     No       &  No   &  No    &   SM     & 95\%  &  77\%   &   Yes   &   Yes    \\ 
5112751&       008002  &   131   &     5112741* &  Yes  &  Yes   &   SM     & 93\%  &  89\%   &   Yes   &   ?      \\ 
5112786&       005003  &   134   &     No       &  No   &  No    &   SM     & 94\%  &  69\%   &   Yes   &   Yes    \\ 
5112880&       002004  &   145   &     No       &  No   &  No    &   SM     & 81\%  &   1\%   &   Yes   &   Yes    \\
5112938&       002006  &   150   &     No       &  Yes  &  Yes   &   SM     & 89\%  &  88\%   &   Yes   &   Yes    \\  
5112948&       005007  &   147   &     No       &  No   &  No    &   SM     & 93\%  &  89\%   &   Yes   &   Yes    \\
5112950&       003005  &   148   &     No       &  No   &  Yes   &   SM     & 95\%  &  92\%   &   Yes   &   Yes    \\  
5112974&       004009  &   151   &     No       &  No   &  Yes   &   SM     & 94\%  &  91\%   &   Yes   &   Yes    \\ 
5113041&       004007  &   153   &     No       &  No   &  No    &   SM     & 94\%  &  26\%   &   Yes   &   Yes    \\ 
5113061&       001014  &   157   &     No       &  No   &  No    &   SM     & 95\%  &  89\%   &   Yes   &   Yes    \\
5113441&       012016  &   185   &     No       &  No   &  No    &   SM     & 89\%  &   0\%   &   Yes   &   Yes    \\ 
5199859&       001016  &    69   &     No       &  No   &  No    &   SM     & 95\%  &  89\%   &   Yes   &   Yes    \\
5200152&       003021  &         &     No       &  No   &  Yes   &   SM     & 94\% &          &   Yes   &   Yes    \\
\tableline											     
\end{tabular}
\tablenotetext{a}{Only targets for which we detect oscillations are listed.}
\tablenotetext{b}{ID, classification, and membership probability from
  radial velocity \citep{Hole09}; SM: single member; BM:
  binary member; BLM: binary likely member; U: Unknown.} 
\tablenotetext{c}{ID and membership probability from proper motion \citep{Sanders72}.}
\tablenotetext{*}{Signal removed from light curve.}
\end{center}}
\end{table*}

\begin{table*}
{\footnotesize
\begin{center}
\caption{Target properties of NGC~6811.\label{tab5}}
\begin{tabular}{cccccccccccc}
\tableline\tableline
Target ID$^a$ & Target ID     & Target ID         & Blend & Blend     & Clump  & Class.     & Membership    & Membership    & Membership        &  Photometric & Seismic  \\
(KIC)         & (Sanders)$^b$ & (Dias et al.)$^c$ & known & potential & star   & (M\&M)$^d$ & (Meibom)$^e$  & (Sanders)$^b$ & (Dias et al.)$^c$ &   member     & member  \\
\multicolumn{1}{c}{[1]} & [2] & [3]               &  [4]  &  [5]      &    [6] &  [7]       &  [8]          &   [9]         &  [10]             &  [11]        &  [12]   \\
\tableline		  	  	    		               			             
9534041       &               &                   &  No   &   No      &  Yes   &            & SLM 74\%      &               &                   &  Yes         &   Yes    \\
9655101       &       95      & TYC3556-00530-1   &  No   &   No      &  Yes   &    SM      & SM  84\%      &  97\%         &   95\%            &  Yes         &   Yes    \\
9655167       &      106      &                   &  No   &   No      &  Yes   &    BM      & BLM 57\%      &  97\%         &                   &  Yes         &   Yes    \\
9716090       &       92      & TYC3556-02356-1   &  No   &   No      &  Yes   &    SM      & SM  78\%      &  94\%         &   95\%            &  Yes         &   Yes    \\
9716522       &      170      & TYC3556-02634-1   &  No   &   No      &  Yes*  &    SM      & SM  79\%      &  97\%         &   97\%            &  Yes         &   Yes    \\
\tableline											     
\end{tabular}
\tablenotetext{a}{Only targets for which we detect oscillations are listed.}
\tablenotetext{b}{ID and membership probability from proper motion \citep{Sanders71}.}
\tablenotetext{c}{Tycho ID and membership probability from proper motion  
  \citep{Dias02}.}
\tablenotetext{d}{Classification from radial velocity measurements 
  \citep{MermilliodMayor90}; SM: single member; BM: binary member.}
\tablenotetext{e}{Classification and membership probability from radial
  velocity measurements (Meibom); SLM: single likely
  member; BLM: binary likely member.}
\tablenotetext{*}{Star appears to be towards the end of He-core burning (Figure~\ref{cmd}).}
\end{center}}
\end{table*}

In addition to the light curve correlations, we identified all
significantly bright stars nearby each target using  
information from the \textit{Kepler Input Catalogue} (\kic), which was
designed to reach down to $Kp\lesssim16.0$. This was just adequate for our purpose. %[maybe use Stetson an Hole instead
%  to see if more complete at these faint stars]  
We identified targets to be potentially affected by
blending if there were stars within five pixels that were at least half as  
bright as the target itself, which we regard a conservative choice given
the results from the above correlation analysis. Potentially blended
targets are listed in Table~\ref{tab1} (column-4), Table~\ref{tab2} (column-5),
and Table~\ref{tab5} (column-5), while the blending stars, their flux 
ratios with respect to the target, and their separations are given in Tables~\ref{tab3}
and \ref{tab4}.  Again, the NGC~6811 targets show no signs of potential
blending. 
Due to slight temporal changes in telescope pointing, the degree of
blending for each target can vary considerably over time for the crowded
cluster fields. We took this into account in our approach to correct the
light curves, which we describe in detail below for
the two clusters affected by blending and other possibly related effects.

\begin{table}
{\scriptsize%\footnotesize
\begin{center}
\caption{Potential blends for NGC~6791 targets.\label{tab3}}
\begin{tabular}{cccc|cccc}
\tableline\tableline
Target ID$^a$ & Blend &   & Sep   & Target ID$^a$ & Blend &   & Sep \\[-2ex]
(KIC)         & (KIC) & \raisebox{1.8ex}{$\frac{\mathrm{Flux_{B}}}{\mathrm{Flux_{T}}}^b$}& (pix) & (KIC )        & (KIC) & \raisebox{1.8ex}{$\frac{\mathrm{Flux_{B}}}{\mathrm{Flux_{T}}}^b$}&  (pix)\\
%Target  &  Blend   & $\frac{\mathrm{Flux_{B}}}{\mathrm{Flux_{T}}}$ & Sep & Target  &  Blend   & $\frac{\mathrm{Flux_{B}}}{\mathrm{Flux_{T}}}$ & Sep  \\
%KICID   &  KICID   &                                               & [pix] & KICID  &  KICID   &  & [pix]  \\
\tableline							       
2297384 & 2297357 & 6.94 & 5.0  & 2437564 & 2437593 & 0.51 & 1.7 \\	 
2435987 & 2435967 & 0.51 & 4.1  & 2437653 & 2437672 & 0.99 & 1.0 \\	 
2436332 & 2436354 & 2.18 & 3.3  &         & 2437641 & 0.67 & 3.2 \\	 
2436458 & 2436455 & 0.53 & 4.8  &         & 2437648 & 5.61 & 2.7 \\	 
2436593 & 2436608 & 2.41 & 1.6  &         & 2437693 & 3.88 & 3.1 \\	 
2436688 & 2436641 & 0.79 & 4.3  &         & 2437706 & 0.79 & 4.1 \\	 
2436759 & 2436750 & 4.28 & 1.0  & 2437781 & 2437775 & 4.24 & 0.8 \\	 
        & 2436746 & 0.51 & 1.0  & 2437804 & 2437803 & 0.69 & 0.8 \\	 
2436814 & 2436879 & 0.85 & 4.8  &         & 2437871 & 0.91 & 4.8 \\	 
        & 2436826 & 1.30 & 0.7  & 2437805 & 2437747 & 0.51 & 4.3 \\	 
2436824 & 2436848 & 0.50 & 2.2  & 2437851 & 2437816 & 8.90 & 3.8 \\	 
2436900 & 2436911 & 0.51 & 1.7  &         & 2437797 & 0.58 & 4.2 \\	 
2436912 & 2436881 & 0.63 & 2.8  & 2437933 & 2437931 & 1.46 & 4.8 \\	 
        & 2436879 & 0.83 & 3.3  &         & 2437928 & 0.55 & 4.4 \\	 
        & 2436897 & 1.52 & 1.4  &         & 2437957 & 1.18 & 4.6 \\	 
        & 2436866 & 2.74 & 2.9  & 2437957 & 2437933 & 0.85 & 4.6 \\	 
2436954 & 2569626 & 0.80 & 2.9  &         & 2437962 & 13.1 & 4.1 \\	 
        & 2436968 & 2.69 & 2.2  &         & 2437972 & 1.13 & 3.3 \\	 
        & 2436973 & 0.98 & 2.2  & 2437965 & 2437999 & 2.80 & 3.9 \\	 
        & 2436944 & 8.77 & 1.2  &         & 2437964 & 0.56 & 2.0 \\	 
        & 2436958 & 5.51 & 4.3  & 2437972 & 2437948 & 0.59 & 2.7 \\	 
        & 2436932 & 0.74 & 2.3  &         & 2437987 & 2.44 & 3.6 \\	 
2437040 & 2437028 & 4.21 & 0.6  &         & 2437962 & 11.6 & 1.1 \\	 
        & 2437022 & 0.63 & 1.5  &         & 2437957 & 0.89 & 3.3 \\	 
2437103 & 2437059 & 1.28 & 4.2  & 2437976 & 2437937 & 3.59 & 3.8 \\	 
        & 2437041 & 0.60 & 4.8  &         & 2437931 & 1.28 & 4.3 \\	 
2437171 & 2437220 & 0.85 & 4.6  &         & 2437932 & 1.87 & 4.3 \\	 
2437240 & 2437220 & 5.90 & 0.9  &         & 2437926 & 0.58 & 4.6 \\	 
        & 2437184 & 0.67 & 3.5  &         & 2437964 & 3.54 & 4.5 \\	 
2437270 & 2437315 & 0.92 & 3.3  & 2437987 & 2437996 & 1.26 & 4.7 \\	 
        & 2437323 & 5.85 & 3.7  &         & 2437962 & 4.76 & 3.2 \\	 
        & 2437299 & 6.47 & 2.2  & 2438038 & 2438078 & 2.81 & 4.0 \\	 
        & 2437348 & 1.92 & 5.0  & 2438051 & 2438073 & 0.79 & 2.2 \\	 
        & 2437234 & 5.18 & 3.1  &         & 2438032 & 0.65 & 5.0 \\	 
        & 2437257 & 3.26 & 2.5  & 2569945 & 2569926 & 5.14 & 2.3 \\	 
        & 2437267 & 2.73 & 2.6  &         & 2569891 & 1.07 & 5.0 \\	 
2437325 & 2437402 & 1.87 & 4.4  &         & 2569925 & 0.85 & 2.7 \\	 
        & 2437313 & 7.21 & 4.2  & 2570094 & 2570091 & 2.15 & 3.3 \\	 
        & 2437329 & 0.64 & 3.7  &         & 2570079 & 2.34 & 1.2 \\	 
        & 2437255 & 0.56 & 4.9  & 2570172 & 2570182 & 0.63 & 2.7 \\	 
2437353 & 2437317 & 0.67 & 4.8  &         & 2570131 & 0.78 & 4.9 \\	 
2437394 & 2437331 & 0.65 & 4.1  & 2570214 & 2570226 & 2.05 & 1.5 \\	 
2437402 & 2437410 & 0.67 & 1.4  & 2570244 & 2570277 & 0.94 & 3.7 \\	 
        & 2437405 & 0.97 & 4.8  & 2570384 & 2570400 & 4.84 & 3.7 \\	 
        & 2437325 & 0.54 & 4.4  &         & 2570370 & 21.4 & 3.5 \\	 
2437488 & 2437437 & 1.26 & 4.5  & 2570518 & 2570524 & 1.08 & 3.7 \\	 
        & 2437429 & 3.06 & 4.7  &         & 2570536 & 2.76 & 3.5 \\   
2437496 & 2437487 & 0.59 & 4.9  &         &         &      &        \\
\tableline 
\end{tabular}
\tablenotetext{a}{Only targets for which we detect oscillations are listed.}
\tablenotetext{b}{Flux ratio between potential blend and target.}
%\tablenotetext{c}{Membership probability from proper motion \citep{Sanders72}.}
\end{center}}
\end{table}

\begin{table}
{\footnotesize
\begin{center}
\caption{Potential blends for NGC~6819 targets.\label{tab4}}
\begin{tabular}{lccc|lccc}
\tableline\tableline
%% Entering 1st row
%& &soft &1 & $-1$ & 1 & 1 & $-1$ & $-1$ & 1 \\[-1ex]
%\raisebox{1.5ex}{Police} & \raisebox{1.5ex}{5}&hard
%& 2 & $-4$ & 4 & 4 & $-2$ & $-4$ & 4 \\[1ex]
Target ID$^a$ & Blend &                                                                & Sep   & Target ID$^a$ & Blend &                                                               & Sep \\[-2ex]
(KIC)         & (KIC) & \raisebox{1.8ex}{$\frac{\mathrm{Flux_{B}}}{\mathrm{Flux_{T}}}^b$}& (pix) & (KIC)         & (KIC) & \raisebox{1.8ex}{$\frac{\mathrm{Flux_{B}}}{\mathrm{Flux_{T}}}^b$}&  (pix)\\
%KICID   &  KICID   &                                               & [pix] & KICID   &  KICID   &                                               & [pix]\\[1ex]
%Target  &  Blend   & $\frac{\mathrm{Flux_{B}}}{\mathrm{Flux_{T}}}$ & Sep   & Target  &  Blend   & $\frac{\mathrm{Flux_{B}}}{\mathrm{Flux_{T}}}$ & Sep \\
%KICID   &  KICID   &                                               & [pix] & KICID   &  KICID   &                                               & [pix]\\
\tableline							       
5024297 &  5024272 & 4.58 & 4.6 & 5024601 &  5024582 & 0.83 & 2.9 \\
        &  5024312 & 0.56 & 2.4 &         &  5112741 & 1.07 & 3.4 \\
5024312 &  5024297 & 1.78 & 2.4 &         &  5112751 & 0.83 & 4.3 \\
        &  5024349 & 3.83 & 4.9 & 5111940 &  5111932 & 2.00 & 2.4 \\
5024405 &  5024410 & 2.55 & 3.2 & 5112467 &  5112445 & 0.69 & 2.4 \\
5024414 &  5024410 & 1.04 & 2.9 &         &  5112478 & 0.80 & 2.4 \\
        &  5024369 & 0.81 & 4.8 &         &  5112491 & 1.04 & 2.9 \\
5024456 &  5024470 & 2.51 & 3.2 & 5112491 &  5112467 & 0.97 & 2.9 \\
5024512 &  5024517 & 1.38 & 3.5 &         &  5112478 & 0.78 & 3.3 \\
        &  5024511 & 2.66 & 3.0 & 5112734 &  5024582 & 1.03 & 3.8 \\
5024517 &  5024511 & 1.93 & 0.9 &         &  5112741 & 1.33 & 2.4 \\
        &  5024512 & 0.73 & 3.5 &         &  5112751 & 1.04 & 3.2 \\
5024582 &  5024601 & 1.21 & 2.9 & 5112751 &  5024582 & 1.00 & 4.5 \\
        &  5112734 & 0.97 & 3.8 &         &  5024601 & 1.21 & 4.3 \\
        &  5112741 & 1.29 & 2.8 &         &  5112734 & 0.97 & 3.2 \\
        &  5112751 & 1.00 & 4.5 &         &  5112741 & 1.29 & 1.7 \\
5024583 &  5024584 & 4.99 & 1.0 & 5112938 &  5112932 & 0.68 & 2.9 \\
\tableline
\end{tabular}
\tablenotetext{a}{Only targets for which we detect oscillations are listed.}
\tablenotetext{b}{Flux ratio between potential blend and target.}
%\tablenotetext{c}{Membership probability from proper motion \citep{Sanders72}.}
\end{center}}
\end{table}

\subsection{NGC~6791}
There is one pair of stars (KIC2569488, 2568916) that show
an abnormally high correlation despite their large separation
($\sim$$\,70\,$pixels) (see Figure~\ref{corr}).  Visual inspection of the light
curves reveals that it arises from a strong trend localized within a
relatively narrow time span.  They are not mutually influenced by a bad
column, neither do they lie within the wings of a really bright star.  From inspection of
the raw images they should not correlate, and there are no indications that
the oscillation signals of the two are affected by each other. 
We therefore leave the light curves uncorrected. 

The Q1 data were not used for KIC2570384, and Q3 was not used for KIC2437488,
both due to apparent blending by short-period eclipsing
binaries. Possible candidates for these blends can be found in
Table~\ref{tab3} (column-2). 
An abnormally strong peak of unknown origin was seen slightly offset from
the main excess envelope in the power spectrum of KIC2569935.  Hence, to
avoid strong bias in the measurement of the asteroseismic parameters, we
fitted and removed a corresponding single sine wave from the data.

\subsection{NGC~6819}
The four stars KIC5024582, 5112734, 5112751, and 5112741 form six
correlation pairs that stand out in Figure~\ref{corr} (encircled).  It
turns out that the increased correlation is caused by the same
large-amplitude contact binary (W UMa variable), KIC5112741.    
Fortunately, we were able to retrieve the pulsation signal to a degree
that allowed the measurement of the seismic parameters used in this paper
(see Sect.~\ref{extract}) except for the W UMa variable itself.  We did
that by applying a high-pass filter to the three affected light curves
using a moving average of two days. % for the first two stars, 
%and one day for the latter.  
The resulting light curves show correlation
coefficients in line with the trend shown by the dashed line in
Figure~\ref{corr}. 
We followed the same approach to remove the variability of what appears to
be a bright M-giant (most likely KIC5024470, Table~\ref{tab4} column-2),
from the light curve of KIC5024456, in this case using a four-day moving
average. 

The Q1 data were not used for KIC5024583 due to the presence of an eclipsing
binary signal.  %strangely enough, its only neighbor, 5024584, shows inverted eclipses
KIC5024268 was also significantly affected by apparent blending
by a $\delta\,$Scuti star in the Q1 and Q2 data. These data were therefore removed
from the light curve.  
Like in one of the NGC~6791 targets, KIC4937770 showed an abnormal peak in
the power spectrum, which we removed by fitting a single sine wave.

\section{Extraction of asteroseismic parameters}\label{extract}
\begin{figure}
\includegraphics{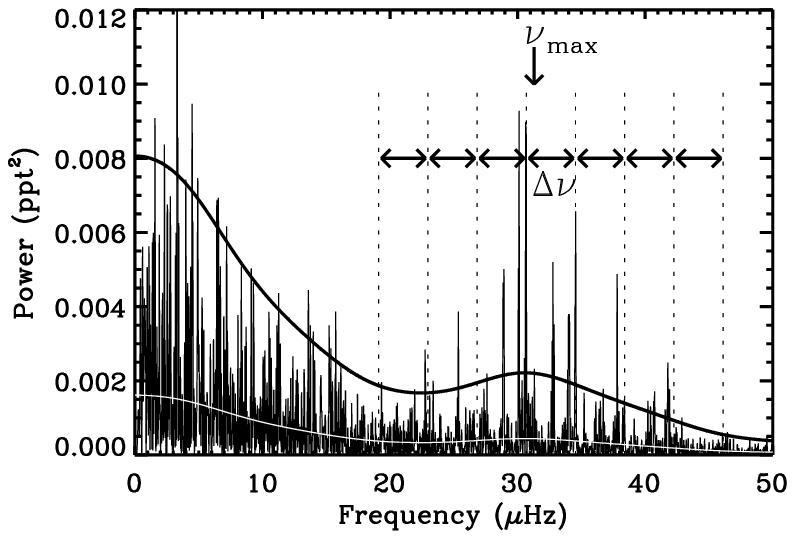}
%\plotone{f1.eps}
\caption{Power spectrum of KIC2436824.  The average large frequency separation,
  \dnu, and the frequency of maximum power, \numax, are indicated with
  arrows.  The equally spaced dashed lines have been positioned to coincide
  with the radial mode near \numax. The solid white curve shows the spectrum
  convolved with a Gaussian function of FWHM $=4$\dnu, which we also show
  after multiplying by five for clarity (thick black curve).
\label{spectrum}} 
\end{figure}

\begin{figure}
\includegraphics{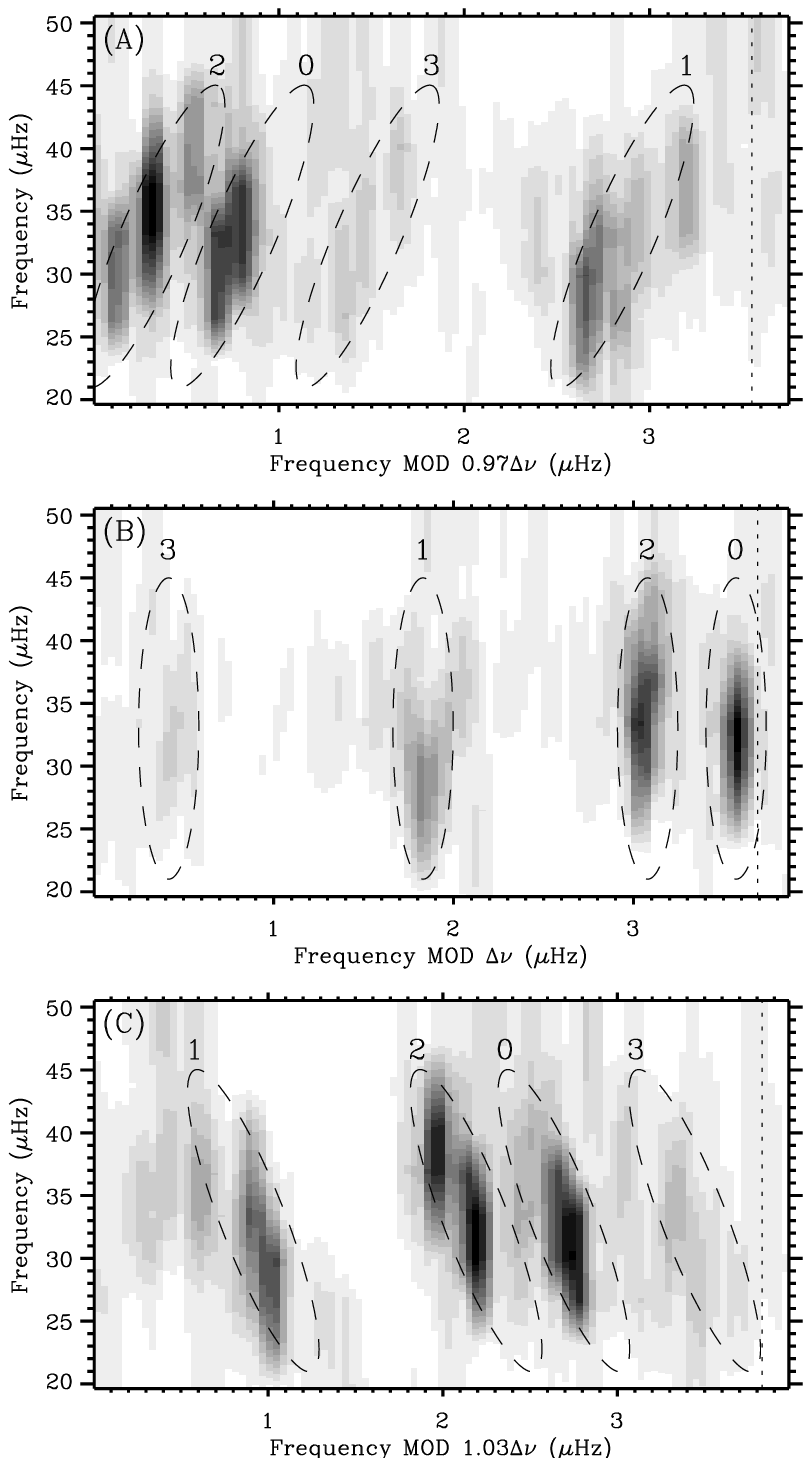}
%\plotone{f1.eps}
\caption{\'Echelle diagrams of the power spectrum of KIC2436824.  The power
  spectrum was smoothed for better visual impression.  Ellipses indicate
  the four ridges formed by oscillation modes of spherical degrees
  $l=0$--3.  The dotted lines mark the expected position of the radial
  ($l=0$) modes (see text).  Each panel shows the \'echelle for slightly
  different choices of the segment width used to divide the power
  spectrum. Panel B shows the best choice of \dnu.  
\label{echelle}} 
\end{figure}

In Figure~\ref{spectrum} we show a typical example of a power spectrum and
indicate for illustration the average large separation, \dnu, and the
frequency of maximum oscillation power,
\numax. 
Values of \dnu~and \numax~ were extracted from the data using each
of the time series analysis pipelines described in 
\citet{Hekker10,Huber09,Kallinger10a,Mathur09a,MosserAppourchaux09}. 
If at least one pipeline detected oscillations, the star was included in
our sample for further investigation.  
The results of all stars were then verified by visual inspection of the power
spectrum and the autocorrelation of the power spectrum.  We further
verified values of \dnu~by forming the so-called \'echelle diagram of the
power spectrum, constructed by dividing the power spectrum into segments,
each \dnu~wide, which were then stacked one above the other.  
To illustrate, Figure~\ref{echelle} shows three examples of the \'echelle
diagram for the same star, each based on a slightly different segment width
(adopted large separation).  
If \dnu~is correct,
the radial oscillation modes form a vertical ridge in the \'echelle, offset from zero
by $\epsilon$ in agreement with recent results on the
$\epsilon$-\dnu~relation of red giants
\citep{Huber10,Mosser10universal,White10}.  
The \'echelle diagram clearly reveals if the adopted large separation is
too small (ridges tilt to the right; panel A) or too large (ridges tilt to
the left; panel C) even by a few percent.
For almost all stars multiple pipelines returned results,
%We received results from multiple pipelines for almost all stars,
and in the vast majority of cases the results agreed within 
a few percent.  %5\%
With such small scatter our conclusions about membership would
essentially be independent on our choice of the final set of results.
We therefore adopted the results from the pipeline that
returned results for most stars, except for the few stars where
%results scattered by more then 5\%.  
\dnu~ was clearly wrong (\'echelle ridges strongly tilted), in which case    
we chose the pipeline results that generated the most vertical ridges in
the \'echelle.
We were able to detect
oscillations in all the red giants in our sample except two stars in
NGC~6819 (KIC5112741 and 5200787) and towards the faint
end of stars in NGC~6791 ($15.8<V<16.8$, $15.5<Kp<16.5$),
for which higher signal-to-noise data, hence longer light curves, will be
required.

\begin{figure}
\includegraphics{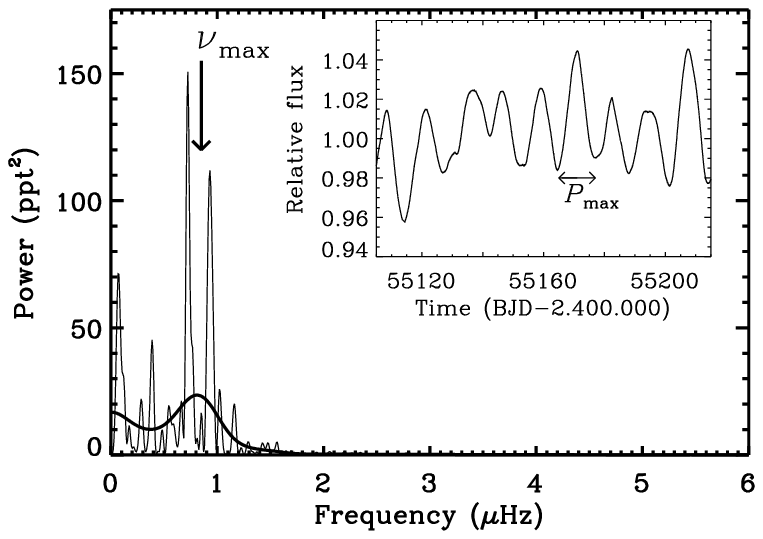}
%\plotone{f1.eps}
\caption{Power spectrum and part of light curve (inset) of one of the long
  period stars where \dnu~could not be reliably determined
  (KIC2437171). The thick curve shows the spectrum convolved with a
  Gaussian function of FWHM $\sim0.4\mu$Hz.  The frequency of maximum
  power, \numax, and corresponding dominant periodicity in the light curve
  are indicated with arrows. 
\label{example}} 
\end{figure}

For some of the most luminous stars, which oscillate at very low
frequencies, \dnu~is quite small and was difficult to determine
reliably with the length of our current dataset. The dominant periodicity,
$P_{\mathrm{max}}$, equivalent to 1/\numax, was however easily detectable
in the Fourier spectrum and even directly in the time series
(Figure~\ref{example}). %, and we were able to measure \numax~down to $\sim0.4\,\mu$Hz. 
The uncertainty in \numax~for these stars is 
relatively large because it was not always possible to correct for the background
granulation signal.

\section{Membership}\label{membership}
Following the approach by \citet{Stello10}, we will use asteroseismic
measurements to categorise our selected stars into asteroseismic cluster
members or likely non-members.  
We note that the length of
the data analysed by \citet{Stello10} only allowed robust measurement of
\numax~for many of 
their targets.  With our current data we can also extract \dnu~for almost
all the target stars, %that show evidence of oscillations, 
which is generally the more precise measurement of the two. 
The former parameter is known to scale with
acoustic cut-off frequency, and hence 
\citep{Brown91,KjeldsenBedding95,Mosser10}: 
\begin{equation}
 \nu_{\mathrm{max}}\simeq\frac{M/M_\odot (T_{\mathrm{eff}}/T_{\mathrm{eff},\odot})^{3.5}}{L/L_\odot}\nu_{\mathrm{max,\odot}},
\label{numax}
\end{equation}
where $T_{\mathrm{eff},\odot}=5777\,$K and \numax$_{,\odot}=3100\,\mu$Hz,
while the latter scales with the square root of the mean density of the star
\citep{Ulrich86,KjeldsenBedding95}: 
\begin{equation}
 \Delta\nu\simeq\frac{(M/M_\odot)^{0.5} (T_{\mathrm{eff}}/T_{\mathrm{eff},\odot})^3}{(L/L_\odot)^{0.75}}\Delta\nu_\odot,
\label{dnu}
\end{equation}
where $\Delta\nu_\odot=135\,\mu$Hz.

\begin{figure}
\includegraphics{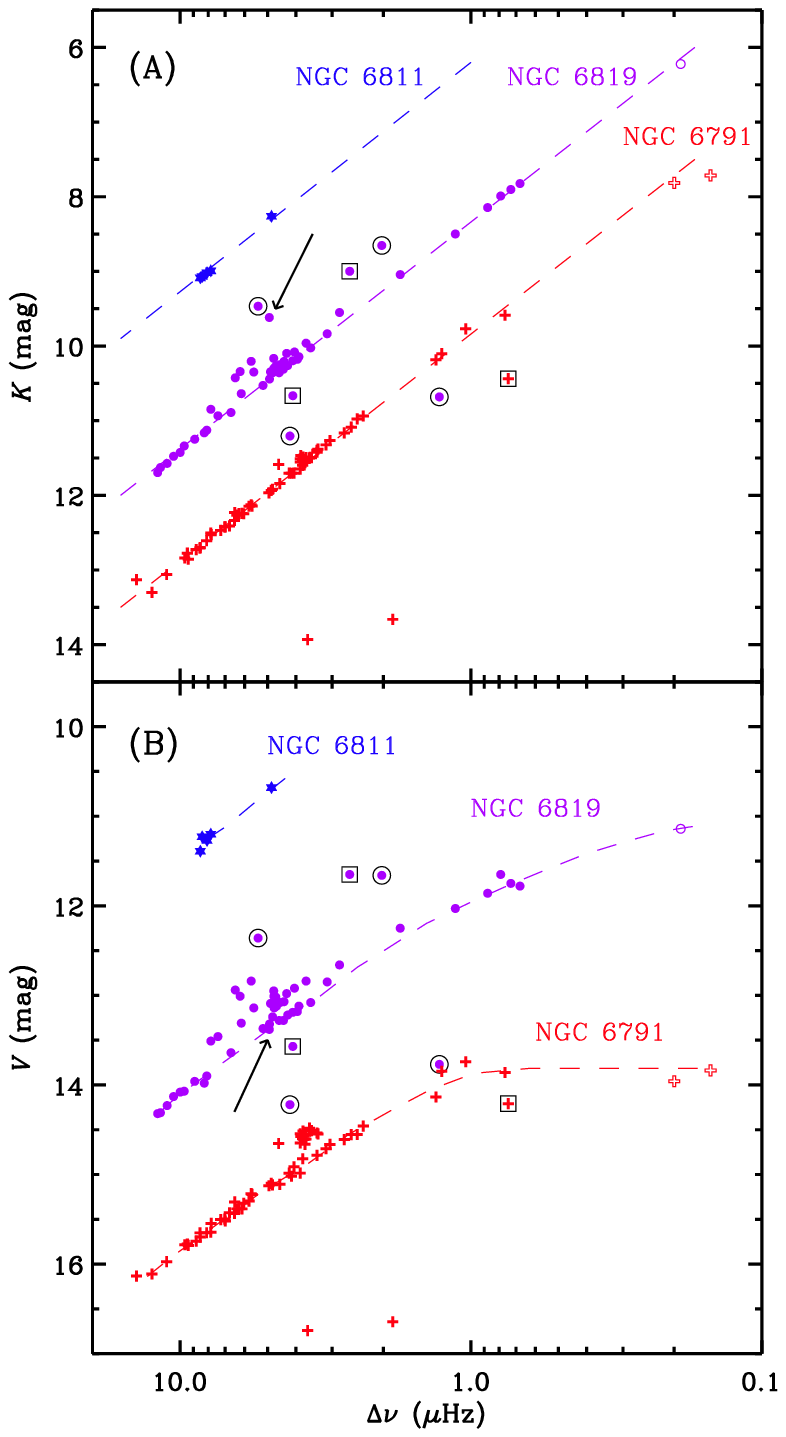}
%\plotone{f1.eps}
\caption{Apparent magnitude versus large separation for NGC~6811 (blue stars),
  NGC~6819 (purple dots), and NGC~6791 (red pluses). Open symbols show
  preliminary results for stars with very low \dnu.  Axes are oriented to
  make the plots resemble a color-magnitude diagram.  Dashed fiducial lines
  are shown to guide the eye.  Arrow marks KIC5024517 (see text).
  Non-members from \citet{Stello10} are encircled.  Newly established
  non-members (this work) are bracketed by squares (see Sect.~\ref{results}). 
\label{dnuVSmag}} 
\end{figure}

Because $M$ and \teff~vary only slightly compared to $L$ within a sample
of cluster red giants, a
tight correlation is expected between both \dnu~or \numax~and the apparent
stellar magnitude, which for cluster members is indicative of luminosity.
By plotting stellar apparent magnitude versus \dnu~or \numax~we indeed see
a tight correlation apart from a few outliers.  To illustrate, we show
2MASS $K$ magnitude \citep{Skrutskie06} versus \dnu~in Figure~\ref{dnuVSmag} (panel A).   
We note that there is no apparent correlation between the crowding value
from \kic~and whether stars follow the expected correlation or not. This
suggests that the \kic~crowding value is not a robust indicator of how much
\dnu~and \numax~are affected by blending. 
This is expected because increased blending does not alter the
oscillation frequencies of the target but only adds extra noise and lowers
the relative amplitude of the oscillations. Only in the rare event where the 
oscillation signal from the blending star is very similar to that of the
target would the measured \dnu~and \numax~be affected.  In
addition, variability from a blending star, such as a binary
companion, could dominate and hence get detected instead of that from the
target. %(or in the case the blending star is a high-amplitude variable)

We made similar plots replacing $K$ band with $J$, $H$, and $V$ to see if
they showed consistent results.  The last is shown in Figure~\ref{dnuVSmag}
(panel B).  The $V$ band clearly shows larger scatter
than the infrared bands due to its stronger sensitivity to differential
interstellar reddening and the slight temperature difference between clump and red-giant-branch
stars of the same mean density.  The bending of the main trend (dashed
lines) is due to strong blanketing affecting the $V$ band measurements for
the cooler stars \citep[see Figure~3 in ][]{Garnavich94}.   
Interestingly, this comparison revealed
that for one star, KIC5024517, the $V$ and $H$ band measurements 
aligned with the expected trend of cluster members, while in $K$ and $J$
bands the star was an outlier (see arrow).  
In addition, there are indications in the power spectrum of excess power 
from two oscillating stars.  The excess located at the highest
frequency (picked up by the time series analysis pipelines) is compatible
with the star being a cluster member if we use the 
$V$ and $H$ band measurements, while the low frequency excess is in
agreement with the $K$ and $J$ band measurements.
This strongly suggests that blending 
(Tables~\ref{tab2} and \ref{tab4}) has affected the standard photometry as
well as the \kepler~light curve.  

\subsection{Estimating \dnu~and \numax}
In the next step, we will estimate the expected \dnu~and \numax~from
solar scaling (Eqs.~\ref{numax} and \ref{dnu}), and compare
them directly with the observations to make inference on cluster
membership.  We do note that because these two parameters are so strongly
correlated \citep{Stello09a,Hekker09,Mosser10}, using both adds little extra
information other than redundancy for the purpose of determining membership. 
%Stellar modelling indicates that any systemetic offset between the scaled and the
%true value varies smoothly as a function of the stellar parameters
%\citep{Stello09a}.   
%The homogeneity of our cluster samples implies that single outliers from
%the expected (scaled) values indicantes that one or more of the stellar
%parameters ($M/M_odot$, $L/L_odot$, or \teff) are not correct, while a
%general offset for the entire ensample of stars could be both due to an
%incorrect mean cluster
%
%Hence, if an offset between the expected (scaled) and observed values is
%detected, it will be in commom for many stars in a homogeneous sample like
%ours,  while single outliers indicate that the scaled value is wrong for
%only that star.
%are  , and single outliers can be intepreted as caused by the observe value. 
%Any star-to-star deviation between expected (scaled) and observed from the 
%expected trend as caused by the observed \dnu~and not the way we
%estimate the expected value from scaling.  
%For our purpose, an average
%offset in \dnu~for the cluster as a whole, is less important.

To estimate the expected \dnu~and \numax~we converted the apparent magnitude
into luminosity, using the cluster distances by
\citet{Basu11}.  For NGC~6811, 
which was not studied by \citet{Basu11}, we adopted a distance modulus of
10.3 mag found by visual isochrone fitting.  The contribution to
the spread in apparent magnitude from the intrinsic depth of the clusters is
% [0.02,0.01,0.005mag for 6811,6819,6791] 
similar to
that from the photometric uncertainty, and is ignored in the following. 
We adopted average cluster reddenings of $E(B-V)=0.16\,$mag (NGC~6791,
\citet{Brogaard11}), $E(B-V)=0.14\,$mag (NGC~6819, \citet{Bragaglia01}), and 
$E(B-V)=0.16\,$mag (NGC~6811, Webda database). Bolometric
corrections were performed using the calibrations by \citet{BessellWood84}
and \citet{Flower96}.

The scatter in the mass of these red-giant-branch stars can be assumed to be low 
(less than 1\% along a standard isochrone, e.g. \citet{Marigo08}).  We
therefore adopted the average red giant mass from \citet{Basu11} for NGC~6791
and NGC~6819.  It should be noted that the assumed common mass might
result in a systematic overestimation of the expected \dnu~for the red
clump stars -- not included in the study by \citet{Basu11} -- if they have
experienced significant mass loss compared to the red-giant-branch stars
\citep{Miglio11}.  We indicate in Table~\ref{tab1} (column-5),
Table~\ref{tab2} (column-6), and Table~\ref{tab5} (column-6) which stars
are clump stars inferred from the color-magnitude diagram (Figure~\ref{cmd}).
For NGC~6811, we used an average mass of $2.35\,$M$_\odot$ derived from
\dnu, \numax, and \teff~by combining Equations~\ref{numax} and \ref{dnu}
\citep{Kallinger10} similar to what was done by \citet{Hekker10a}.  
Finally, we emphasize that the absolute values adopted for the average cluster
properties are of less importance since we are looking only to distinguish 
stars that deviate from the average trend.  Hence, it is important to
take into account the relative \teff~and bolometric corrections of each
star.   

\begin{figure}
\includegraphics{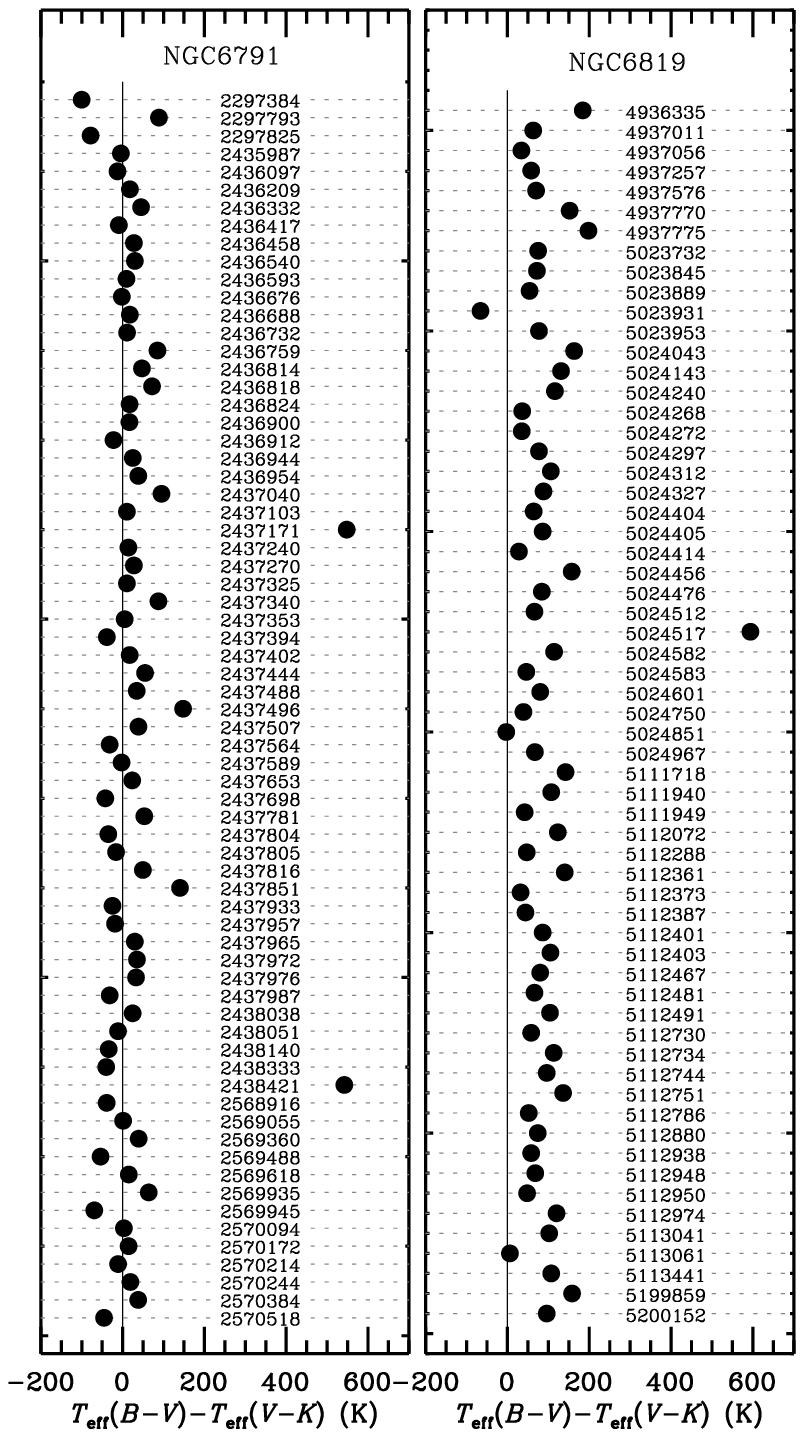}
\centering{\includegraphics{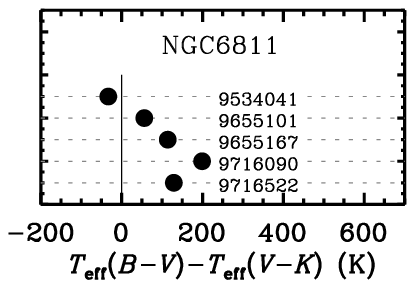}}
%\plotone{f1.eps}
\caption{Difference between temperatures derived from the $B-V$ and $V-K$
  color indices. 
\label{tempcompare}} 
\end{figure}

To obtain \teff~we transformed the $V-K$ color index using the calibrations by
\citet{RamirezMelendez05}. We estimate the uncertainty in \teff~to be
$\sim100\,$K, which includes contributions from the
photometry, color-temperature calibration, and reddening (\citealt[see~][~for
further details]{Hekker10a}). 
As a double check we compared our \teff~with those derived from $B-V$.  The
results are shown in Figure~\ref{tempcompare}. 
The offset seen for NGC~6819 shows a slight dependency with color and is
most likely due to calibration errors in the standard photometry. 

The $V-K$ index is genrally the better temperature proxy for cool red
giants, it is less sensitive to metalicity, and its temperature calibration
show lower scatter than for $B-V$.  We therefore choose $V-K$. 
The two outliers, KIC2437171 and KIC2438421, are much cooler than they
appear in $B-V$ \citep{Garnavich94}, supporting that $V-K$ is the prefered
color index.
However, for KIC5024517 -- the outlier in NGC~6819 -- we recall that the
$K$-band measurement is probably affected by blending
(Figure~\ref{dnuVSmag}), suggesting $V-K$ might not be the best
temperature indicator in this case.

\subsection{Results on the observed-to-expected ratio}\label{results}
\begin{figure}
\includegraphics{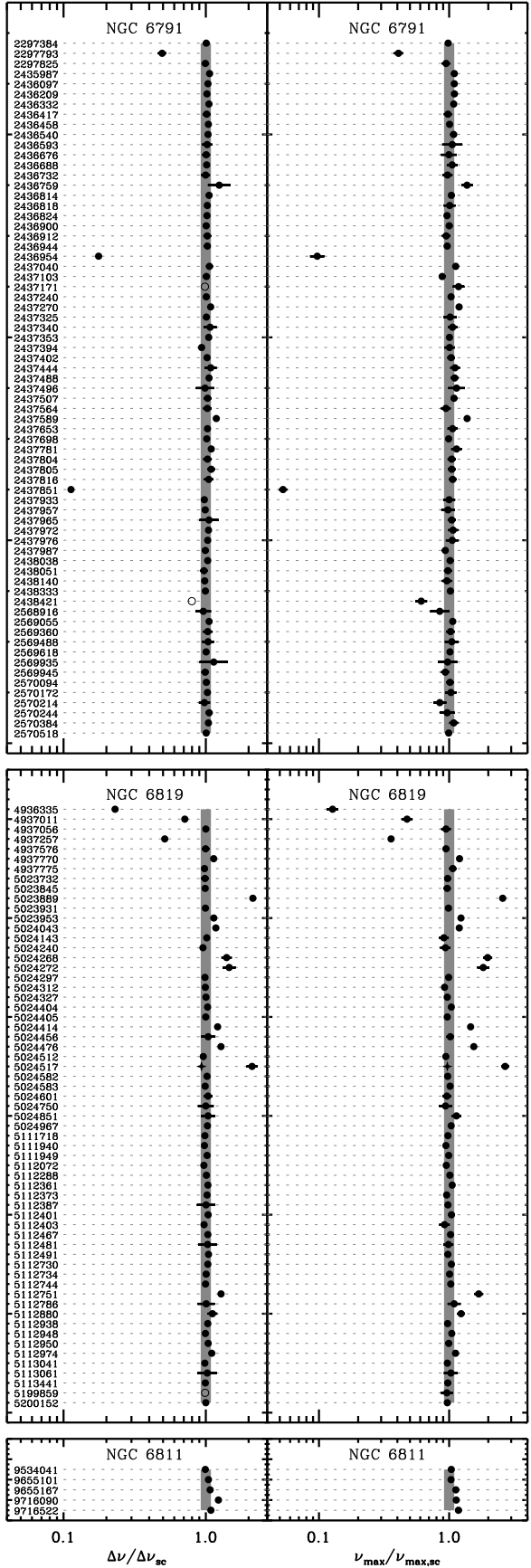}
%\plotone{f1.eps}
%\vspace{-1.cm}
\caption{Ratio between measured and solar-scaled \dnu~(left) and \numax~(right)
  for targets in NGC~6791 (top), NGC~6819 (middle), and NGC~6811
  (bottom). The 1-$\sigma$ region of the expected ratio is shown in gray.
  Open symbols show preliminary results for stars with very low \dnu.  Star
  symbols show results using the $B-V$ temperature (see text for details).
\label{ratio}} 
\end{figure}

The observed-to-expected ratios for \dnu~and \numax~are shown in
Figure~\ref{ratio}.  The expected ratio is 1.0, with an uncertainty
slightly below 10\% (1-$\sigma$ region marked in gray), which is dominated
by the uncertainty in \teff.  The size of the uncertainty underpins
  that ignoring the expected spread in mass of $\lesssim1\%$ is sound.  For our 
purpose the absolute value 
of the average ratio is not important but rather the deviation of single stars
from the ensemble.  However, it turns out that the majority of stars fall close
to 1.0, which shows that any possible systematic errors in the expected 
(scaled) \dnu~and \numax, caused by offsets in the calibration of the scaling
relations or inaccurate adopted cluster parameters, have cancelled out.
Apart from a few clear outliers we see generally little
scatter, which indicates that blending is less of an issue than suggested
by the large number of potential blends, particularly for NGC~6791
(Table~\ref{tab1}, column-4). 
We note that we did not find any general trends between on the one hand
stellar brightness or color and the other hand the deviation of \dnu~and
\numax~from the cluster average.  Nor did we detect a difference in
  the observed-to-expected ratios between stars on the red gaint branch and
  the red clump; hence supporting that any possible mass loss, which is
  expected to occur predominantly near the tip of the red giant branch,
  is insignificant compared to the uncertainty in the ratios
  plotted in Fig.~\ref{ratio}.
However, stars that do clearly deviate need to be carefully assessed 
before we draw any conclusions about their cluster membership. 
We will discuss each cluster in turn. 

In NGC~6791 there are three outliers in \dnu~(KIC2297793, 2436954, 2437851)
and an additional borderline case in \numax~(KIC2438421).  The latter is
one of the stars with very low \dnu~and \numax~for which accurate
uncertainties were difficult to determine and we therfore do not make a final
conclusion on its membership.  Of the others, KIC2436954 and 2437851 are the two
faintest stars ($V\simeq16.1$, $Kp\simeq16.4$) for which
oscillations have been detected.  The former shows strong evidence
of blending, while the latter is potentially blended.  
%Three additional
%stars show unexpected low \numax~(KIC2297793, 2437171, 2438421), 
%all of which also show evidence of a large frequency separation in agreement with the
%\numax~measurement.  However, due to their intrinsically high luminosity (low
%\numax), we see only modes in two or three consecutive orders, which are
%barely resolved (an example is shown in Figure~\ref{example}). While we
%do not claim detection of \dnu~in these stars, we
%regard the \numax~measurements as solid.  
%Of the latter two, KIC2437171 and 2438421, only KIC2437171 is known to be
%blended (Table~\ref{tab1} column-3), but both show abnormal differences
%between \teff~derived from $V-K$ and $B-V$ indicating they might in fact
%both be blended or that they have been mismatched. 
%We note that adopting the $B-V$ temperature does not resolve
%the discrepancy.  
This leaves KIC2297793 as the only star where we can not
explain the results as potentially due to blending.  The fast stage of
evolution of this very luminous star (upper 
red giant or asymptotic giant branch assuming it is a cluster member) means that a potential binary
companion would presumably be much fainter, making it unlikely that the
companion affects our measurements significantly (both seismic as well as
the standard photometry).  We therefore conclude the most likely
explanation is that the star is not a cluster member.
For some of the targets we have membership probabilities from 
radial velocity (Meibom \& Platais, priv. comm. (2010) and 
\citet{Garnavich94}), and from both radial velocity and metallicity
\citep{WortheyJowett03,Origlia06,Carraro06,Gratton06},  which we list in
Table~\ref{tab1} (column-6--8), 
while column-9 lists our seismic membership results.  Our results on
KIC2297793 reaffirm that of \citet{Garnavich94} who's ambiguous 'no?'
designation was chosen because the star's radial velocity was significantly
different from the cluster average despite moving in the same direction as the
cluster as opposed to the bulk of the field.
Finally, it is noticeable that quite a few seismic members --
all initially selected as photometric members -- are assigned low probability
membership from the radial velocity survey by (Meibom \& Platais).  We
speculate that this could be due to binary companions, which we will address
using seismology in a forthcomming paper.

NGC~6819 shows a few more outliers than NGC~6791 despite the stars being
brighter and less affected by blending (compare Tables~\ref{tab3} and
~\ref{tab4}).  Our results confirm all four seismic non-members
identified by \citet{Stello10} KIC4936335, 4937257, 5023889, and 5024272.
In addition, we can identify two new seismic non-members, KIC4937011 and
5024268. 
All other apparently discrepant stars could be explained by potential
blending or binarity (KIC5024414, 5024476, 5024517, and 5112751) affecting
both the seismic measurements and the temperature estimates (spectroscopic
binaries are listed in Table~\ref{tab2} column-7).  In fact,
adopting the temperature from $B-V$ for KIC5024517 makes it agree quite
well (star symbol).  We see the opposite in KIC5023931, which is a known
binary where the $V-K$ temperature and the detected oscillations agree with
membership, while it would appear as a non-member if we adopt the $B-V$
temperature. 
%The last star that appears to be a non-member (KIC5199859), has a very low
%\numax.  Its power spectrum was therefore analysed 
%manually and the uncertainty was estimated by extrapolating from the
%stars with the lowest \numax~detected by the pipelines.  This might be an
%underestimate, and we conclude it is too soon to make firm conclusion
%about the membership of this star. 
For comparison we list membership probabilities from radial velocity, and
proper motion in Table~\ref{tab2} (column-8 and 9), as well as photometric
membership alongside the identified seismic membership (column-10 and 11).
In four cases stars appear to be both seismic and photometric non-members.
Three stars (KIC5112880, 5113041, and 5113441) have very low proper motion
probability contradicting the results from radial velocity, photometry, and
seismology.

Finally, all five stars selected in NGC~6811 show seismic signals that
agree with them being cluster members (Table~\ref{tab5}, column-12). 
%KIC9716090 is the only star showing
%a slight, however not significant, offset from the expected value, but only
%so if the $V-K$ temperature is used.  
Our results agree well with those inferred from radial velocity measurements
(column-7 and 8), proper motion (column-9 and 10), and photometric
membership (column-11).

\section{Discussion and conclusion}
We have demonstrated that cluster membership determined from seismology show 
advantages over more orthodox methods, and hence offers important
complementary information to that of kinematic and photometric measurements.  
In our asteroseismic investigation we implicitly assumed a standard
evolution history for the cluster stars when estimating the seismic
parameters.  Any exotic stars, such as remnants of strong dynamic
interactions, would therefore appear as non-members under
this assumption.  While a kinematic study does not assume a standard stellar
evolution, it can assign spurious field stars as members and vice versa if
the space velocity of the cluster is not clearly distinct from that of the
field.  The asteroseismic determination, however, is insensitive to that, as
it essentially separates non-members from members by revealing stars that
are at a different distance than the ensemble mean. 

From almost a year of \kepler~data, we were able to measure global seismic
properties of over a hundred red giant stars in three open clusters
allowing inference to be made on the cluster membership of each star.
Among our list, comprising likely members determined from photometric and
kinematic surveys, we found three new non-members and confirmed
the four previously identified seismic non-members by \citet{Stello10}.
We found more seismic non-members in NGC~6819 despite this cluster having
fewer members than NGC~6791. This could indicate issues with obtaining a
clean sample of 
members purely from the kinematic properties of NGC~6819. % arising from
%random alignment of the space velocity of some field stars with that of
%the cluster.   
However, we note that a probably significant contribution to the
difference in number of identified non-members in these two clusters 
comes from the different selection criteria adopted for each cluster.  
%For
%NGC~6791 the selection was based on a relatively strict photometric
%criterion, which we had to apply due to the lack of a comprehensive
%kinematic study of this cluster (Sect~\ref{selection}). 

Finally, we highlighted that the presence of binary stars and blends
needs careful investigation to avoid misinterpretation of the
seismic results as well as the auxiliary standard photometry.  In some
cases the light curves revealed seismic signal from more than one star,
which could be used to identify and to some extent disentangle signals from
blends and binaries.

Future \kepler~data of the so-called cluster super stamps will give access
to most stars in NGC~6791 and NGC~6819, providing more comprehensive
assessment of the effects from blending, and enable us to make unbiased
selections of the cluster stars, which will further extend the asteroseismic 
analyses of these clusters.  
In addition, we will get short cadence data (1-minute sampling) of 
selected stars in NGC~6819 which will allow us to probe the interiors of 
the less evolved turn-off and subgiant stars. 

%COR sent after frist draft (see below for sent after second draft)
%2297793 Dnu = 0.76 muHz ; numax = 3.9 muHz
%2437171 Dnu = 0.23 muHz ; numax = 0.75 muHz
%2438421 Dnu = 0.27 muHz ; numax = 0.84 muHz
%5199859 Dnu ~ 0.20 muHz ; numax ~ 0.77 muHz 
%         COR             SYD                            CAN
%KIC      Dnu     numax   Dnu           numax            Dnu           numax
%2297793  0.76    3.9                   3.993 0.072      0.744 0.0239  3.917 0.159
%2437171  0.23    0.78    0.200 0.007   0.857 0.019  
%2438421  0.27    0.84    0.150 0.005   0.400 0.009
%5199859  0.17    0.64    0.190 0.007   0.710 0.016   
%    Uncertaintly ~10%

\acknowledgments
Funding for this Discovery mission is provided by NASA's Science Mission Directorate.
The authors would like to thank the entire {\it Kepler} team without whom this
investigation would not have been possible.
DS acknowledges support from the Australian Research Council.
This project has been supported by the `Lend\"ulet'
program of the Hungarian Academy of Sciences and
the Hungarian OTKA grant K83790 and MB08C 81013.
SH acknowledges financial support from the UK Science and Technology
Facilities Council (STFC). KB acknowledges financial support from the
Carlsberg Foundation. NCAR is supported by the National Science Foundation.
RS thanks the support of the J\'anos Bolyai Research Scholarship.

\end{document}